\documentclass[twocolumn,floatfix,groupedaddress,superscriptaddress,aps,prb,showpacs]{revtex4-1}

\usepackage{graphicx}
\usepackage{amsmath}
\usepackage{amssymb}
\usepackage{amsfonts}
\usepackage{dsfont}
\usepackage{dcolumn}
\usepackage{bbm}
\usepackage{float}
\usepackage{subfigure}
\usepackage{mathtools}
\usepackage{braket}
\usepackage{placeins}
\usepackage[unicode]{hyperref}
\usepackage[nameinlink]{cleveref}
\Crefname{section}{Sec.}{Secs.}
\Crefname{equation}{Eq.}{Eqs.}
\Crefname{figure}{Fig.}{Figs.}
\Crefname{tabular}{Tab.}{Tabs.}

\usepackage{bbold} % For \mathbb{1}
\usepackage{nicefrac}
\usepackage{tikz}

\newcommand{\bes}{\begin{subequations}}
\newcommand{\ees}{\end{subequations}}

\newcommand{\Liouville}[1]{\mathcal{L}\!\left(#1\right)}
\newcommand{\LiouvilleSub}[2]{\mathcal{L}_{#1}\!\left(#2\right)}
\newcommand{\matrixsymbol}[1]{\mathbf{#1}}
\renewcommand\vec{\mathbf}
\newcommand{\normfactor}[0]{\mathcal{N}}
\newcommand{\scalarpr}[2]{\left(#1\middle|#2\right)}

\newcommand{\erz}[1]{f_{#1}^\dagger}
\newcommand{\erzVer}[1]{f_{#1}^{(\dagger)}} 
\newcommand{\ver}[1]{f_{#1}^{\vphantom{\dagger}}}

\newcommand{\bzo}[1]{\widehat{n}_{#1}}
\newcommand{\bzoMod}[1]{\widehat{n}_{#1}^{\vphantom{\dagger}}}

\newcommand\leftidx[3]{
  {\vphantom{#2}}#1#2#3
}

\newcommand{\conjugateTransposeSuperscript}[3]{\leftidx{}{#1}{^{#3\dagger}_{#2}}}

\newcommand{\normalOrder}[1]{{:}\mkern.4mu#1\mkern.6mu{:}}

\newcommand{\wone}[1]{w_1^\dagger(#1)}
\newcommand{\wtwo}[3]{w_2^\dagger(#1,#2,#3)}
\newcommand{\wthree}[3]{w_3^\dagger(#1,#2,#3)}
\newcommand{\wfour}[3]{w_4^\dagger(#1,#2,#3)}
\newcommand{\wfive}[3]{w_5^\dagger(#1,#2,#3)}
\newcommand{\wsix}[3]{w_6^\dagger(#1,#2,#3)}
\newcommand{\wseven}[3]{w_7^\dagger(#1,#2,#3)}
\newcommand{\weight}[3]{w_8^\dagger(#1,#2,#3)}
\newcommand{\wnine}[3]{w_9^\dagger(#1,#2,#3)}

\newcommand{\threeBasis}{$3$-basis}
\newcommand{\threePlusBasis}{$3^{+}$-basis}

\newcommand{\dv}[1]{\frac{\text{d}}{\text{d}#1}}
\newcommand{\Tr}{\text{Tr}}
\newcommand{\dd}{\text{d}}
\newcommand{\expval}[1]{\langle #1\rangle}
\newcommand{\comm}[2]{[#1,#2]}
\newcommand{\ev}[2]{\langle #2| #1|#2\rangle}

\begin{document}

\title{Strong quenches in the one-dimensional Fermi-Hubbard model}

\author{Philip Bleicker}
\email{philip.bleicker@tu-dortmund.de}
\affiliation{Lehrstuhl f\"{u}r Theoretische Physik I, 
Technische Universit\"{a}t Dortmund,
 Otto-Hahn Stra\ss{}e 4, 44221 Dortmund, Germany}

\author{G\"otz S.\ Uhrig}
\email{goetz.uhrig@tu-dortmund.de}
\affiliation{Lehrstuhl f\"{u}r Theoretische Physik I, 
Technische Universit\"{a}t Dortmund,
 Otto-Hahn Stra\ss{}e 4, 44221 Dortmund, Germany}

\date{\textrm{\today}}

\begin{abstract} 
The one-dimensional Fermi-Hubbard model is used as testbed for strong global parameter quenches.
With the aid of iterated equations of motion in combination with a suitable scalar product for
operators we describe the dynamics and the long-term behavior in particular of the system
after interaction quenches. 
This becomes possible because the employed approximation allows for oscillatory dynamics avoiding
spurious divergences. The infinite-time behavior is captured by an analytical approach
based on stationary phases; no numerical averages over long times need to be computed. 
We study the most relevant frequencies in the dynamics after the quench and find that the 
local interaction $U$ as well as the band width $W$ dominate.
In contrast to former studies a crossover instead of a sharp dynamical transition depending 
on the strength of the quench is identified. 
For weak quenches the band width is more important while for strong quenches the local
interaction $U$ dominates.
\end{abstract}

%05.70.Ln	Nonequilibrium and irreversible thermodynamics
%67.85.−d	Ultracold gases, trapped gases
%71.10.Fd	Lattice fermion models (Hubbard model, etc.)
%71.10.Pm 	Fermions in reduced dimensions

\maketitle

\section{Introduction}
\label{s:introduction}

Systems far away from thermal equilibrium give rise to fascinating properties and thus have been
a source of inspiration for finding both highly non-linear material characteristics and 
studying the evolution of strong correlations. Unfortunately, most of these studies had to remain gedanken experiments for long time with no feasible experimental realization. But in
 recent years the research in non-equilibrium physics gained steam mainly due to remarkable
 experimental progress which renders a dedicated preparation and observation of 
non-equilibrium phenomena possible.

The creation and precise tuning of optical lattices to confine ultra-cold atomic gases
\cite{Anderson1998,Bloch2005,Trotzky2011} form the basis for experimentally analyzing 
former purely theoretical Hamilton operators
\cite{Greiner2002,Goldman2016}. Moreover, femtosecond spectroscopy and pump-probe experiments 
allow one to gain insight into the evolution of ultrafast correlations in solid state physics
\cite{Axt2004,Morawetz2004,Perfetti2006}. Various invasive and non-invasive imaging processes
 have been proposed to perform in-depth studies of quantum states. The use of 
Bragg spectroscopy and time-of-flight experiments \cite{Stoferle2004}, 
\textit{in situ} techniques with fluorescence \cite{Sherson2010}, matter-wave scattering
\cite{Sanders2010,Mayer2014}, optical cavities
\cite{Mekhov2007}, or Dicke superradiance \cite{TenBrinke2015} is possible for this purpose.

Groundbreaking experimental progress induces an urgent demand for corresponding 
theoretical descriptions and powerful tool kits for non-equilibrium phenomena. 
Systems away from equilibrium are usually in highly excited states so that the occurring processes are spread over wide scales of energy and hence of time. For this reason, 
common techniques from equilibrium physics are often not applicable. 
The hugely varying time scales are illustrated by the relaxation times of doublons (double occupancies) in  Mott insulators which are shown to be different from intrinsic time scales 
of the system by orders of magnitude \cite{Strohmaier2010,lenar13}. 
Especially the enormous number of excitations in the system makes the 
usual theoretical description in terms of a few dressed quasi-particles
\cite{Mattuck1976} insufficient.

A suitable method to prepare a system out of equilibrium in order to study the
ensuing dynamics is to quench the system, i.e., to change its parameters abruptly.
This approach has been used very frequently, e.g., in one-dimensional Bose-Hubbard systems 
for quenches across quantum phase transitions both theoretically \cite{Lauchli2008} and experimentally \cite{Chen2011} or to observe propagations of thermal correlations by 
coherently splitting a one-dimensional Bose gas into two separate parts \cite{Langen2013}.

There is a number of theoretical tools to describe the non-equilibrium time evolution. 
Special systems allow for analytic treatments 
\cite{cazal06,barth08,Calabrese2011,Calabrese2012a,calab12b,Caux2013}.
Exact diagonalization \cite{Rigol2009} is very flexible, but limited in the maximum size 
of the system. Non-equilibrium dynamical mean field theory \cite{Eckstein2009,Aoki2014}, or 
perturbative expansions in the inverse coordination number \cite{Navez2010,Krutitsky2014} 
work both best for infinite or large dimensions. The time-dependent density matrix 
renormalization group \cite{white04a,Daley2004} is most powerful in one dimension and 
simulations by quantum Monte Carlo\cite{Batrouni2005,Goth2012} rely on detailed balance 
so that they are inherently designed for equilibrium configurations. Variational Gutzwiller approaches \cite{Schiro2010,Schiro2011} provide an analytical approach which captures
quantum fluctuations only partly; variational quantum Monte Carlo is a very powerful
technique, but computationally very expensive \cite{ido15}. 
Continuous unitary transformations \cite{Moeckel2008,Moeckel2009,Sabio2010} 
have so far been employed in leading order in $U$ only.

In spite of the above list of approaches the need persists to improve and extend the theoretical tool box, in particular for the understanding of
the temporal evolution on very long time scales including the infinite-time averages.
These infinite-time averages are crucial because they characterize the stationary state
to which the quenched system evolves. Relevant issues are the question whether 
these stationary states are thermal Gibbs ensembles and on which time scales they
are reached.
 
In the present paper we aim at enlarging the theoretical tool set with the stationary states
in mind by starting from iterated Heisenberg equations of motion 
\cite{Uhrig2009}. Former studies based on this approach were able to accurately describe only
short times due to spurious diverging dynamics induced by the necessary approximations
\cite{Hamerla2013,Hamerla2013a,Hamerla2014,Kalthoff2017}. It was realized that these
divergences are due to non-unitary evolution generated by inappropriate truncations 
\cite{Kalthoff2017}. A general remedy for systems with finite local Hilbert spaces
was sketched based on a suitable scalar product for operators.
Here, we realize this idea for a fermionic model
showing that the time evolutions computed in this way
indeed avoid spurious divergences so that long time behavior can be discussed
and expectation values in the stationary state are accessible.
For a spin model, namely the central spin model, the approach advocated here
has been applied successfully already \cite{Rohrig2017}.

Since we aim at a proof-of-principle illustration of the promising ideas
we choose a relatively simple one-dimensional Fermi-Hubbard model as testbed.
We stress that its integrability \cite{Lieb1968,Essler2005} is no prerequisite for the
applicability of our approach. On the contrary, 
we stress that the approach is applicable to arbitrary finite dimensions. 
Using iterated equations of motion the thermodynamic limit can be treated 
as well.

The setup of this article is as follows. In \Cref{s:model}, the model is introduced and the general quench protocol is explained briefly. Thereafter, \Cref{s:method} summarizes the concept of iterated equations of motion and introduces the scalar product which preserves unitarity in
 terms of operators. A so far unexplored analytical way to compute infinite-time averages is outlined and realized for three important observables. The ensuing results
 are presented in \Cref{s:observables}. The results are summarized in 
\Cref{s:summary} while \Cref{s:outlook} provides an outlook.

\section{Model} 
\label{s:model}

\begin{figure}[htb]
  {\centering
  \includegraphics[width=.85\columnwidth,clip]{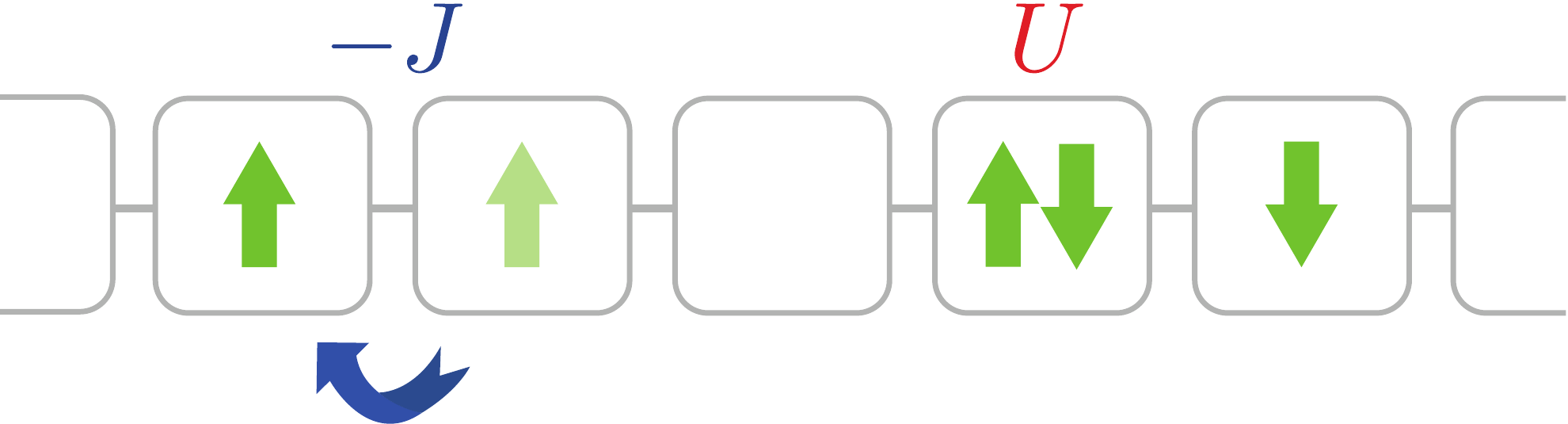}
  \caption{(Color online) Sketch of the one-dimensional Fermi-Hubbard model in real space. Electron hopping is determined by the matrix element $J$, double occupancy of a site costs an additional energy $U$.
  \label{fig:fh_model}}}
\end{figure}

The Fermi-Hubbard model consists of tight-binding electrons with strongly screened Coulomb 
interaction \cite{Hubbard1963,Kanamori1963,gutzw63}. We consider
nearest-neighbor hopping and a completely local repulsion with one band in one dimension, see
\Cref{fig:fh_model}, with the Hamilton operator
\begin{equation}
	\label{eq:fh_model}
	H = H_0 + H_\text{int} = -J \sum_{\substack{\left<i, j \right> \sigma}} \erz{i\sigma} \ver{j\sigma} + U \sum_{i\sigma} \bzo{{i\uparrow}}\bzo{{i\downarrow}}
\end{equation}
where $J$ denotes the hopping matrix element and $U$ the local interaction, i.e., the 
energy cost of a double occupation. The Fermi-Hubbard model is often used to describe 
electronic properties of condensed matter systems with narrow energy bands, 
metal-insulator-transitions or even high-temperature superconductors. 
A natural energy scale of its kinetic hopping part is given by the band width $W$, which
is given by $2zJ$  with the coordination number $z$ for bipartite lattices. 
Henceforth, we use $J$ as energy unit; concomitantly all times are measured 
in units of $\nicefrac{1}{J}$.

To generate a non-equilibrium state a quantum quench is used. Initially, the system is prepared
in an eigenstate of $H_0$; for simplicity, we use the Fermi sea $\ket{\text{FS}}$.
After the quench the time evolution is governed by a different Hamiltonian, i.e., by
the full Fermi-Hubbard Hamiltonian $H_0 + H_\text{int}$ including the on-site interaction. 
Thus, the explicit time dependence of the Hamiltonian is expressed by 
$H_\text{Q}(t) = H_0 + \theta(t) H_\text{int}$
where $\theta(t)$ is the Heaviside function.
The state of the system deviates noticeably from $\ket{\text{FS}}$ for times $t>0$. Since
the quench in $H_\text{Q}(t)$ changes an overall system parameter which influences all sites
it is called a global quench. Global quenches are widely considered
\cite{manma07,Moeckel2008,Moeckel2009,Manmana2009,Barmettler2009,Calabrese2011,Calabrese2012a,calab12b,Caux2013}.

As mentioned in the introduction, we do not exploit or consider the
integrability of the above model but use it as simple testbed to illustrate
the advocated theoretical approach. Our goal is to compute long-time behavior
including infinite-time averages in a systematically controlled way.

\section{Method} 
\label{s:method}

\subsection{Dynamics}
\label{ss:dynamics}

In order to deduce the time dependence of operators we resort to the iterated equations of
 motion approach \cite{Uhrig2009,Hamerla2013,Hamerla2013a,Hamerla2014,Kalthoff2017}, 
a brief summary of which is given in the first part of this section. The second part is dedicated to the necessary modifications of the method which warrant 
a unitary time evolution on the operator level.

Let us consider an arbitrary operator in the Heisenberg picture
\begin{equation}
	\label{eq:op_expansion}
	A(t)=\sum_i h_i(t)A_i
\end{equation}
whose time dependence is completely contained in the complex prefactors $h_i(t)$ where the 
constant operators $A_i$ form a suitable operator basis. Throughout this article $\hbar$ 
is set to unity for simplicity. At this point, only the linear independence of the $A_i$
is necessary to make \eqref{eq:op_expansion} well-defined.
Let there be no explicit time dependence of the Hamiltonian 
such that the Heisenberg equation of motion simplifies to
\begin{equation}
	\label{eq:h_eom}
	\dv{t} A(t) = i\comm{H(t)}{A(t)} =: i \Liouville{A(t)}
\end{equation}
with the Liouville superoperator $\Liouville{\cdot}$. 
Inserting \eqref{eq:op_expansion} into \eqref{eq:h_eom} leads to
\bes
\begin{align}
	\label{eq:eom_op_expansion}
	\dv{t}A(t) &=i\Liouville{A(t)}\\
	& = i\sum_i h_i(t)\Liouville{A_i}.
\end{align}
\ees
It is convenient to define linear expansions for all operators $\Liouville{A_i}$
\begin{equation}
	\label{eq:liouville_auf_op_expansion}
	\Liouville{A_i} := \sum_j M_{ji} A_j
\end{equation}
leading to the Liouvillian matrix $\matrixsymbol{M}$, also called dynamic matrix.  
It is convenient to
combine the  time dependent prefactors $h_i(t)$ to a vector $\vec{h}(t)$.
Its dynamics  is governed by
\begin{equation}
\label{eq:matrix_dgl}
	\dv{t}\vec{h}(t) = i \matrixsymbol{M} \vec{h}(t).
\end{equation}

In the Schr\"{o}dinger picture, the time evolution of the states is determined by the 
unitary time evolution operator which reads $e^{-iHt}$ for a constant Hamiltonian that has no 
explicit dependence on $t$. Hence, all solutions are superpositions of oscillatory terms
whose frequencies are given by the eigenenergies.
Phenomena such as dephasing or relaxation only occur as superpositions of infinitely
many terms with continuously distributed frequencies.
Mapping the time dependence onto operators and thus switching to the Heisenberg picture 
does not alter the outcome so that again the temporal evolutions is given by
superpositions of oscillatory terms.

Inappropriate approximations, however, will leave us with a dynamic matrix 
$\matrixsymbol{M}$ which has complex eigenvalues $z_i$. Then, exponential behavior arises
and if the matrix itself is real the complex eigenvalues occur in pairs 
$z_\pm = R \pm iI$ with real part $R$ and imaginary part $I$. 
Clearly, one of them induces an exponential divergence in time.
Even if the eigenvalues stay real,  the matrix may not be diagonalizable, but
of Jordan normal form, so that divergences occur which are of power law type.
Such divergencies can only be avoided if we can guarantee that $\matrixsymbol{M}$
is diagonalizable with real eigenvalues. 

A sufficient, though not necessary, condition
to guarantee that $\matrixsymbol{M}$
is diagonalizable with real eigenvalues is to show that it is Hermitian \cite{Kalthoff2017}.
If $\matrixsymbol{M}=\matrixsymbol{M}^\dagger$ is given one knows that
the matrix always possesses real-valued eigenvalues $\lambda_j$ with equal 
algebraic and geometric multiplicity such that a general solution to 
\Cref{eq:matrix_dgl} can always be written as
\begin{equation}
	\label{eq:matrix_dgl_solution}
	\vec{h}(t)=\sum_{j=1}^f \alpha_j e^{i\lambda_j t}\vec{v}_j
\end{equation}
with the corresponding eigenvectors $\vec{v}_j$ and coefficients $\alpha_j$ chosen according to the given initial conditions. Clearly, the general solution is the superposition of
oscillatory terms.

How can one be sure that $\matrixsymbol{M}$ is Hermitian in an approximate treatment?
In order to compute the matrix easily, it is convenient to use an orthonormal 
operator basis $\{A_i\}$ (ONOB) so that 
\begin{equation}
	\label{eq:matrix_elements_onb}
	M_{ji}=\scalarpr{A_j}{\Liouville{A_i}}
\end{equation}
holds. Hermitecity of $\matrixsymbol{M}$ means $M_{ji} = M_{ij}^*$. It follows
from $\Liouville{\cdot}$ being self-adjoint. Whether this is the case or not
depends on the choice of the scalar product. Hence, its choice is crucial. 

Following Ref.\ \onlinecite{Kalthoff2017}, we define the 
operator scalar product for two linear operators $A$ and $B$ 
defined on a Hilbert space $\mathcal{H}$ as
\begin{equation}
	\label{eq:frobenius_scalar_product}
	\scalarpr{A}{B} := \normfactor \Tr(A^\dagger B)~\text{with}~\normfactor
	:=\frac{1}{\Tr(\mathbb{1})}.
\end{equation}
In so doing, we implicitly assume the (local) Hilbert space to be finite, i.e., 
$\operatorname{dim}(\mathcal{H}) < \infty$. This clearly holds for all spin systems, 
fermionic systems such as the Fermi-Hubbard model in question, or models with both spin 
and fermionic degrees of freedom. Bosonic degrees of freedom have to be excluded.

From a physical perspective, the above defined scalar product equals 
the high-temperature limit $T\to\infty$ of the thermal expectation value
\bes
	\label{eq:high_temperature_limit_frobenius}
\begin{align}
	\scalarpr{A}{B} &= \lim_{T\to\infty} \expval{A^\dagger B}
	\\
	&= \lim_{T\to\infty} \Tr \left(\rho A^\dagger B\right)
\end{align}
\ees
in the canonical ensemble for a density matrix $\rho = \nicefrac{e^{-\beta H}}{Z}$ with the partition sum $Z=\Tr\left(e^{-\beta H}\right)$, Hamiltonian $H$ and the 
inverse temperature $\beta \ge 0$. Accordingly, the considered system is maximally disordered and each state of the Hilbert space 
is equally likely due to $\lim_{T\to\infty} \rho \propto \mathbb{1}$.

The self-adjointness of $\Liouville{\cdot}$ stems from the invariance of the trace 
under cyclic permutations and allows us to show that $\matrixsymbol{M}$
is Hermitian
\bes
\begin{align}
\label{eq:Hermitian_Liouville_matrix}
	M_{ji} &= \normfactor \Tr(A_j^\dagger \comm{H}{A_i})
	\\
	&= \normfactor \Tr(A_j^\dagger (H A_i-A_i H))
	\\
	&= \normfactor \Tr(\comm{H}{A_j}^\dagger A_i)
	\\
	&= \scalarpr{A_i}{\Liouville{A_j}}^* \\
	&= M_{ij}^*.
\end{align}
\ees
Note that there are also other scalar products with the property that
the Liouvillian is self-adjoint with respect to them, for instance
the expression on the right hand side of \eqref{eq:high_temperature_limit_frobenius}
at finite temperatures if the ensemble is taken with respect to the
Hamiltonian after the quench. But such a scalar product would in practice
require detailed knowledge of the Hamiltonian after the quench, for instance
its diagonalization. Hence, the choice \eqref{eq:frobenius_scalar_product}
is very advantageous in the sense that it is easy to use and generally applicable 
not requiring any particular knowledge of the system. Another asset is
that truly time dependent Hamiltonian $H(t)$ with $\partial_t H\ne 0$ for almost
all times can also be tackled with the choice \eqref{eq:frobenius_scalar_product}
while more sophisticated choices depending on the actual Hamilton operator
would yield time dependent scalar products.

If one follows the advocated strategy it is clear that the time evolution
in the Heisenberg picture is unitary in the sense that the operator scalar
product remains constant, i.e., for any $A$ and $B$ we have
\begin{equation}
(A(t)|B(t)) = (A(0)|B(0)).
\end{equation}
This is what is meant by unitarity on the operator level.
It can be formally expressed by 
\begin{equation}
A(t) = \mathcal{U}(t) A.
\end{equation}
implying also that $\expval{A(t)}$ consists of the sum of oscillatory terms.

But it does \emph{not} imply unitarity on the level of \emph{states} as one knows
it from text book quantum mechanics where the unitary solution $U(t)$ of
\begin{equation}
i\dv{t}U(t) = H(t)U(t)
\end{equation}
implies that the scalar product of two arbitrary states $\ket{a}$
and $\ket{b}$ stays constant
\bes
\begin{align}
\bra{a(t)}b(t)\rangle &= \bra{a(0)}U^\dag(t)U(t)b(0)\rangle
\\
&= \bra{a(0)}b(0)\rangle.
\end{align}
\ees
Note that unitarity of the states \emph{implies}
unitarity of operators if we define for any operator $A$
\begin{equation}
\label{eq:unitarity1}
\mathcal{U}A:= U^\dag A U.
\end{equation}
Then we can conclude
\bes
\begin{align}
(A(t)|B(t)) &= \mathcal{N} \Tr(U^\dag A^\dag U U^\dag B U)
\\
&= \mathcal{N} \Tr(A^\dag  B )
\\
&= (A(0)|B(0)).
\end{align}
\ees

The inverse is not true: unitarity on the level of operators
does not imply unitarity on the level of states.
Inspection of \eqref{eq:unitarity1} makes this plausible
because the left hand side does not impose any particular
structure on $\mathcal{U}$ while the right hand side 
does. 

An even clearer piece of evidence results from the 
inspection of fermionic anticommutators $\{\erz{i},\ver{j}\}=\delta_{ij}$.
Clearly, they stay constant under unitary transformations of
the states, i.e., for $\erz{i}(t)=U^\dag \erz{i} U$ and 
$\ver{j}(t)=U^\dag \ver{j} U$ we have
\bes
\label{eq:anticommut1}
\begin{align}
\{\erz{i}(t),\ver{j}(t)\} & = U^\dag (\erz{i} U U^\dag \ver{j}+ \ver{j} U U^\dag \erz{i}\}) U
\\
&= \delta_{ij}.
\end{align}
\ees
Note that the above identity implies a large number of scalar equations
for any given pair $i,j$
because it is an \emph{operator} identity. For an $F$ dimensional Hilbert space
it amounts up to $F^2$ scalar equations.

Generically, \eqref{eq:anticommut1} does not hold for a unitary transformation on
the operator level because such a transformation only guarantees the conservation of 
operator scalar products, i.e., for $\erz{i}(t)=\mathcal{U} \erz{i}$ and 
$\ver{j}(t)= \mathcal{U}\ver{j}$ we have
\bes
\label{eq:anticommut2}
\begin{align}
\mathcal{N}\Tr\left(\{\erz{i}(t),\ver{j}(t)\}\right) & = \mathcal{N}
\Tr\left(\{\erz{i},\ver{j}\}\right)
\\
&= \delta_{ij}.
\end{align}
\ees
Although the above relation looks similar to \eqref{eq:anticommut1}
it only represents a \emph{scalar} identity for any given pair $i,j$.
In other words, a unitary transformation on operator level is much less
restricted than a unitary transformation for states. 
Still, the unitary time evolution on operator level ensures that the solutions
are superpositions of oscillatory terms without any divergencies.

\subsection{Time-dependent expectation values}
\label{ss:time-averages}

We intend to consider three observables. The first observable is the 
expectation value of the local particle number 
\begin{equation}
	n_{i\sigma}(t):=\expval{\bzo{i\sigma}(t)} = 
	\ev{\erz{i\sigma}(t)\ver{i\sigma}(t)}{\text{FS}}
\end{equation}
which describes the expectation value of the
 number of particles present at lattice site $i$ with spin $\sigma$ at time $t$. 
The expectation value is taken with respect to the ground state of the non-interacting model, i.e., the Fermi sea, because we choose this state as the initial state before
the quench.

The second observable is the momentum distribution
\begin{equation}
	n_{k\sigma}(t):=\expval{\bzo{k\sigma}(t)} = 
	\ev{\erz{k\sigma}(t)\ver{k\sigma}(t)}{\text{FS}}
\end{equation}
where $k$ is the wave vector and $\erz{k\sigma}(t)$ and $\ver{k\sigma}(t)$
the Fourier transforms of the creation and annihilation operators in real space.
Further details are given in \Cref{ss:momentum_distribution} where the corresponding
results are shown.

The third observable is the jump at the Fermi surface $\Delta n(t)$ defined by
\begin{equation}
	\label{eq:jump}
	\Delta n(t) := \lim_{\mathclap{k \to k_\text{F}^-}} n_{k\sigma}(t) - 
	\lim_{\mathclap{k \to k_\text{F}^+}} n_{k\sigma}(t)
\end{equation}
for the Fermi wave vector $k_\text{F}$.
The limits in \eqref{eq:jump} are meant as one-sided limits and denoted by 
negative and positive superscripts, respectively.

Since all observables involve one-particle operators they can be easily computed from the 
time evolution of the elementary fermionic creation and annihilation operators 
$\erzVer{i\uparrow}(t)$. The most general ansatz reads
\begin{equation}
	\label{eq:ansatz_erz}
	\erz{i\uparrow}(t) = P_i^\dagger + \left[P^\dagger \left(P^\dagger H^\dagger \right)\right]_i 
	+ \ldots
\end{equation}
for the creation operator \cite{Uhrig2009,Hamerla2013}. Here, $P$ stands for a
general particle creation operator or a linear superposition of several of them and
$(PH)$ stand for a general combination of creation and annihilation operator (hole creation)
or a linear superposition of them. The subscript $i$ stands for the site on the lattice
around which  the operator superposition is located. The terms left out and 
only indicated by the dots are terms comprising two and more particle-hole pairs.

Concretely, the superposition $P_i$ of particle creation operators reads
\begin{equation}
	\label{eq:ansatz_superposition}
	P_i^\dagger := \sum_{\delta \lessapprox v_\mathrm{max}t} \sum_\sigma p_{i\pm\delta,\sigma}^*(t) \erz{i\pm\delta,\sigma}
\end{equation}
with scalar coefficients $p_{j,\sigma}$.
Here, we used that the superposition spreads around its origin at site $i$ only 
at a finite velocity $v_\text{max}$ so that for a given time $t$ 
significant contributions occur only in a restricted cone 
$[i-v_\text{max}t, i+v_\text{max}t]$ according to the Lieb-Robinson bound
\cite{Lieb1972}. Contributions outside of the cone given by the group velocity $v_\text{max}$
are exponentially suppressed. 
If long-ranged interactions are present no such linear cones
occur \cite{Jurcevic2014a,Hauke2013,Eisert2013,Junemann2013}.
But this is not the case we are considering here.

In the following, we use the general representation 
\begin{equation}
\erz{\ell\uparrow} = \sum_m h^{(\ell)*}_m(t) A_m^\dagger
\end{equation}
 with $A_m$ chosen according to \Cref{eq:ansatz_erz}. The superscript indicates the initial lattice site at which the particle is put into the system. Both the occupation number operator 
and the jump are bilinear expressions in the prefactors $h^{(\ell)}_n$. Each term 
 $h^{(\ell)*}_m(t) h^{(j)}_n(t)$ is multiplied with the expectation value of the 
corresponding operators, i.e., with 
\begin{equation}
\label{eq:Amatrix-def}
A_{mn}=\langle A_m^\dag A_n\rangle.
\end{equation}
These values are taken as matrix elements of the matrix 
$\matrixsymbol{A}$.
Hence, the time dependence of a general bilinear term is given by
\begin{equation}
\label{eq:general}
\expval{\erz{\ell\sigma}\ver{j\sigma}}(t)
= \vec{h}^{(\ell)\dag}(t) \matrixsymbol{A} \vec{h}^{(j)}(t).
\end{equation}
Note that this evaluation can be numerically demanding because it requires
to sum twice over the index of the ONOB. It turns out that often
the numerical solution of the differential equations \eqref{eq:matrix_dgl}
is not the limiting factor, but the actual computation of \eqref{eq:general}.

The expectation values $A_{mn}$ are computed in the initial state.
Hence, it depends on the initial state which values enter and how
easy or complicated it is to determine $\matrixsymbol{A}$.
In this paper, we choose the Fermi sea as initial state so that
the expectation values $A_{mn}$ can be easily computed 
by factorizing them using Wick's theorem \cite{Wick1950}. 

Computing the jump at the Fermi surface requires some additional
considerations.
Since the Fermi jump is a rescaled Heaviside-like discontinuity it is 
exclusively determined by terms proportional to $\nicefrac{1}{r}$, i.e., by terms which 
have the longest range in real space. A Fourier transform of all these terms yields 
$\Delta n(t)$. The terms in $\matrixsymbol{A}$ with the longest-range  
are those which result from the single-particle excitations \emph{relative}
to the Fermi sea \cite{Uhrig2009,Hamerla2013}. To extract these terms we 
proceed by normal-ordering the general ansatz \eqref{eq:ansatz_erz} for
the fermionic creation operator
\begin{equation}
  \label{eq:normal_ordered_op_expansion}
  \erz{i\uparrow}(t) = \underbrace{\sum_{m}^N H_m^{(i)*}(t)\,
	\normalOrder{\erz{m\uparrow}}}_{\substack{\text{one-particle}\\
	\text{contributions}}} +
	\normalOrder{\left[P^\dagger\left(P^\dagger H^\dagger\right)\right]_i\!} 
	\,+ \ldots \ .
\end{equation}
Only the first term relates to single-particle excitations and thus matters for the jump
$\Delta n$. Here, additional superscripts clarify which initial conditions are used, i.e., they are used to denote the lattice site at which a particle is inserted at time $t=0$. For general observables superscripts are omitted for brevity.

We highlight that the approach advocated here consists of two steps, in contrast
to what has been realized previously \cite{Hamerla2013,Hamerla2013a,Hamerla2014}.
In the first step, an ONOB is used to describe the
evolution of operators in the Heisenberg picture. No normal-ordering enters
at this stage differing from what has been done before. 
Only in the second step, we normal-order the operators of the ONOB
in order to distill the single-particle part relevant for the jump
at the Fermi level.

The concrete procedure runs as follows.
Consider the annihilation operator and let
\begin{equation}
  \label{eq:basis_mapping}
  H^{(0)}_n(t) = \sum_j t_{nj} h^{(0)}_j(t)
\end{equation}
where $h^{(0)}_j(t)$ is the time-dependent prefactor of the operator $A_j$ and
$t_{nj}\in\mathbb{C}$ quantifies to which extent the normal-ordering
of the operator $A_j$ contributes to the single-particle operator 
$\ver{n\sigma}=\normalOrder{\ver{n\sigma}}$, see also \eqref{eq:normal_ordered_op_expansion}.
Hence, the factors $t_{nj}$ are two-point expectation values or
sums of products of them according to Wick's theorem.
For instance, $t_{5j}=2^{3/2}\expval{n_3-1/2}$ if $A_j=2^{3/2}\ver{5\uparrow} 
(\erz{3\downarrow}\ver{3\downarrow}-1/2)$.
Hence, the coefficients $t_{nj}$ represent the effect of normal-ordering,
given that an arbitrary operator basis has been chosen before.

It is an advantage of the Fermi jump that it can be computed 
from the coefficients $H_n$. Each of these coefficients
requires only a single sum over the ONOB whereas other observables
require a double sum over the ONOB which is of large
dimensionality. We once again stress that the actual integration of the differential equation
\eqref{eq:matrix_dgl} does not represent the bottleneck generically, but
the final evaluation of the expression \eqref{eq:general}.

Once the one-particle prefactors $H_n(t)$ are determined, the jump $\Delta n(t)$
can be computed either by
\begin{equation}
\label{eq:jump-time} %rechts \frac{1}{N^2}
\Delta n(t) = \sum_{m,n}^N H^{(0)*}_m(t)H^{(0)}_n(t) e^{i k_\mathrm{F} (m-n)}
\end{equation}
or by first computing the Fourier series
 \cite{Uhrig2009,Hamerla2013,Hamerla2013a} 
\bes
\begin{align}
H^{(0)}_k(t) &=\sum_n H^{(0)}_n(t) \exp(-ikn)
\end{align}
and then taking the square of its absolute value
\begin{align}
\Delta n(t) &= |H^{(0)}_k(t)|^2.
\end{align}
\ees

\subsection{Infinite-time averages}
\label{ss:long_term}

We focus here on the temporal evoluation for long times in particular.
The long-term behavior of observables contains information about whether and to which 
extent a quenched system retains information about its initial state. Moreover, it provides
evidence if and how the system approaches stationary states. In particular, the averages over
infinitely long time intervals provide information about the expectation values of
the stationary state.

In this section we analytically derive such averages for observables. While many approaches
require to numerically compute the temporal evolution in order to
finally average over it our method directly addresses the averaged quantities. 
The observable average for $t\to\infty$ can be computed based on the knowledge of the initial correlation at time $t=0$ without the calculation of any time dependence. 
Of course, it is also \emph{possible} to average a computed temporal evolution over
suitably chosen time intervals, for instance in order to address the same
quantities as measured in experiment.

We define the infinite-time average of an observable by
\begin{equation}
	\label{eq:long_term_behaviour}
	O_\infty:=\lim_{t\to\infty} \frac{1}{t}\int_0^t\dd t' \expval{O(t')}.
\end{equation}
This definition is particularly helpful in situations without a well-defined infinite-time 
limit $\lim_{t\to\infty}\expval{O(t)}$ due to non-vanishing oscillatory contributions. Calculating the infinite-term average \eqref{eq:long_term_behaviour} can be performed in a fully analytical approach provided that the constraints of \Cref{ss:dynamics} are met.

Due to the equality of Heisenberg and Schr\"{o}dinger picture at $t=0$ 
the prefactors $\alpha_j$ in \eqref{eq:matrix_dgl_solution} 
are determined. For simplicity, we include these initial conditions in 
scaled eigenvectors by defining
\begin{equation}
\vec{\overline{v}}_j:=\alpha_j \vec{v}_j.
\end{equation}
Note that only the $\alpha_j$ depend on different initial conditions.
In the following, $\overline{v}_{p,q}$ denotes the $q$-th component of the 
scaled eigenvector $\vec{\overline{v}}_p$ and thus describes the contribution to 
$h_q(t)$ in \Cref{eq:op_expansion}. 

Inspecting the temporal evolution of the prefactors $\vec{h}(t)$
in \eqref{eq:matrix_dgl_solution} the oscillatory contributions are shown to vanish in the infinite-time averages which implies that only
the terms with stationary phases contribute to them
\cite{Fioretto2010}. For the infinite-time average we obtain
\bes
  \label{eq:long_term_product_final}
  \begin{alignat}{2}
    &\lim_{t\to\infty} \frac{1}{t}\int_0^t\dd t' h^{*}_m(t') h_n(t') 
		\\&= \sum_{i,j} \overline{v}_{i,m}^{\,*} \overline{v}_{j,n} \underbrace{\lim_{t\to\infty}
		\frac{1}{t}\int_0^t\dd t' e^{i(\lambda_j-\lambda_i)t'}}_{=\,\delta_{\lambda_i,\lambda_j}} 
		\\
      &= \sum_{\substack{i,j\\\lambda_i=\lambda_j}} \overline{v}_{i,m}^{\,*} \overline{v}_{j,n}.
  \end{alignat}
\ees
Consequently, only contributions located within the same subspace spanned by the
eigenvectors of the same eigenvalue do not vanish in the long run. 

Applying \eqref{eq:long_term_product_final} to the infinite-time average 
of the local particle number operator the following relation holds
\begin{subequations}
  \label{eq:n_infty_2}
  \begin{alignat}{2}
  n_\infty &= \lim_{t\to\infty} \frac{1}{t}\int_0^t\dd t' \sum_{m,n} 
	h^{(0)*}_m(t') h^{(0)}_n(t') 	\expval{A_m^\dagger A_n} 
	\\
  		   &= \sum_{\substack{i,j\\
				\lambda_i=\lambda_j}} \conjugateTransposeSuperscript{\overline{\vec{v}}}{i}{(0)}
	\matrixsymbol{A}\overline{\vec{v}}^{(0)}_j
  \end{alignat}
\end{subequations}
where we used the expectation value matrix $\matrixsymbol{A}$
defined by its matrix elements in \eqref{eq:Amatrix-def}.

Finally, we turn to the Fermi jump. Its general time dependence
is given by \Cref{eq:jump-time}. 
Averaging over infinite time using \eqref{eq:long_term_product_final} we obtain
\begin{subequations}
  \begin{alignat}{2}
  &\lim_{t\to\infty} \int_0^t\dd t' H_m^{(0)*}(t') H_n^{(0)\vphantom{*}}(t') \\
  &= \sum_{p,q} t_{mp}^*\, t_{nq}^{\vphantom{*}} \lim_{t\to\infty} \int_0^t\dd t' h_{p}^{(0)*}(t') 
	h_{q}^{(0)}(t') \\
  \label{eq:HmHn_infty}
   &= \sum_{\mathclap{\substack{p,q,i,j\\\lambda_i=\lambda_j}}} t_{mp}^* t_{nq}^{\vphantom{*}} \overline{v}_{i,p}^{(0)*} \overline{v}_{j,q}^{(0)} \\
   &= \sum_{\mathclap{\substack{i,j\\\lambda_i=\lambda_j}}} 
	\conjugateTransposeSuperscript{\overline{\vec{v}}}{i}{(0)}
	\matrixsymbol{T^{mn}}\overline{\vec{v}}^{(0)}_j
  \end{alignat}
\end{subequations}
with the usually highly sparse transformation matrix 
$\matrixsymbol{T^{mn}}$ defined by its matrix elements
\begin{equation}
T_{pq}^{mn} := t_{mp}^* t_{nq}^{\vphantom{*}}.
\end{equation}
Consequently, the infinite-time average of the jump can be
expressed very concisely by
\begin{equation}
  \label{eq:Delta_n_infty}
  \Delta n_\infty = \sum_{m,n}^N \sum_{\mathclap{\substack{i,j\\\lambda_i=
	\lambda_j}}} \conjugateTransposeSuperscript{\overline{\vec{v}}}{i}{(0)}
	\matrixsymbol{T^{mn}}\overline{\vec{v}}^{(0)}_j \,e^{i k_\mathrm{F} (m-n)}.
\end{equation}
This concludes the general part of the advocated approach.
We stress again, that the above method allows us to address
directly the expectation values of the stationary state assumed
at infinite times.

\subsection{Lanczos algorithm}

The calculations in \Cref{ss:long_term} are analytical ones, but their
evaluation finally involves numerics. Thus, it is instructive
 to discuss numerical implications. 
For the infinite-time averages eigenvalues and eigenvectors are needed so that
a full diagonalization of the Liouville matrix is required.
Its run time scales like $\mathcal{O}\left(F^3\right)$ 
where $F$ is the dimension of the ONOB. This may become
a very large number so reductions of the matrix dimensions are of interest.
An efficient technique for achieving this goal
must eliminate less needed directions leading to a reduced matrix 
$\matrixsymbol{\widetilde{M}} \in \mathbb{C}^{f \times f}~\text{with}~f < F$ 
with modified eigenvectors which span a smaller subspace 
$\widetilde{S}$ than the original 
$S=\mathrm{span}\left(\left\{\vec{v}_j\right\}\right)$. 

Accordingly, not all former vectors $\vec{r} \in S$ are still element of
$\widetilde{S}$. In order to fulfill the initial condition truncation has to ensure that
$\vec{h}(0) \in \widetilde{S}$ holds. The Lanczos algorithm \cite{lancz38} fulfills
these requirements. It is a special case 
of the Arnoldi iteration \cite{Arnoldi1951} for Hermitian matrices. 
Starting from the initial condition vector $\vec{s}:= \vec{h}(0)$ 
it gradually constructs the $f$-dimensional Krylov space of operators
\begin{equation}
  \label{eq:krylov_space} 
  \mathcal{K}^{f}\!\left(\vec{s}\right)=\mathrm{span}\left(\vec{s},\,\matrixsymbol{M}\vec{s},\,\matrixsymbol{M}^2\vec{s},\ldots,\,\matrixsymbol{M}^{f-1}\vec{s}\right).
\end{equation}
The time required to diagonalize the smaller tridiagonal Liouville matrix 
$\matrixsymbol{\widetilde{M}}$ in this Krylov space 
can be significantly reduced. The precise gain in run time depends on the 
ratio of $f$ to $F$.

\subsection{Operator monomials}

For arbitrary $0<\nicefrac{J}{U}<\infty$ each application of the Liouville operator, i.e., 
each commutation with $H$, is prone to create operator monomials which were not yet
in the considered basis. In the following discussion all time-dependent prefactors 
are omitted for brevity, but one should keep in mind that each operator monomial
comprises also such a prefactor. The shorthand 
$\LiouvilleSub{p}{\cdot}$ stands for a commutation with $H_p$ where $p=0$ or $p=\text{int}$. 
The set of lattice sites in real space
where an operator monomial has a non-trivial effect, i.e., has
a local operator different from the identity, is called the corresponding
cluster. Often, the terms ``operator monomial'' and ``cluster''
are used interchangeably if one focusses on the relevant sites.

\begin{figure}[htb]
  {\centering
  \includegraphics[width=\columnwidth,clip]{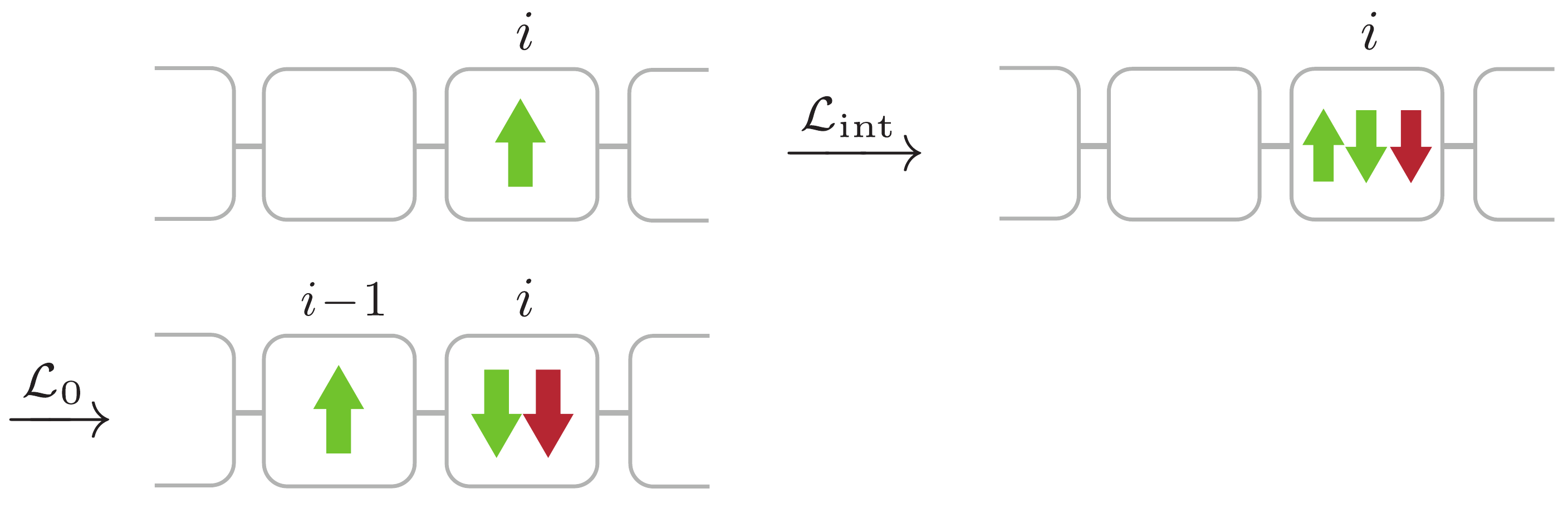}
  \caption{(Color online) Combined effect of hopping and interaction: Green (red) colored 
	arrows stand for particle (hole) creation operators, the spin direction is described 
	by the arrow orientation.
  \label{fig:hubbard_1D_kinAndInt}}}
\end{figure}

\Cref{fig:hubbard_1D_kinAndInt} graphically illustrates 
the effects of both hopping and interaction on operator monomials.
The hopping part $H_0$ moves 
operators through the lattice and the interaction part generates monomials with 
increasing numbers of operators. Consequently, no finite operator basis 
is closed under iterated commutation with $H_0$ and $H_\text{int}$.
Generically, the number of sites involved proliferates upon commutation, i.e.,
the clusters continuously grow.

Not all operators contribute to the dynamics to the same extent
depending on the parameter regime. 
Exemplarily, for the limit of strong on-site repulsion, i.e., 
$\nicefrac{J}{U}\ll1$, hopping represents a small perturbation and 
plays a minor role since nearly local processes dominate the time evolution of the system. We focus on this particular regime and refrain from including physically strongly suppressed 
operators in the ONOB. This means that monomials generated by many hopping processes
are neglected.

Periodic boundary conditions for a lattice of $N$ sites are used.  We choose a 
fixed ONOB constructed with respect to locality and orthonormality in the sense of
the scalar product \eqref{eq:frobenius_scalar_product}. 
We start from local creation operators, apply the Liouville operator and
project the resulting expression onto the subspace spanned by operators in the chosen
ONOB. Roughly, a larger ONOB with more operators is expected to provide better results
\cite{Kalthoff2017}.

Concretely, we apply $\LiouvilleSub{0}{\erz{i\uparrow}} \propto \erz{i\pm1\uparrow}$ and 
$\LiouvilleSub{\mathrm{int}}{\erz{i\uparrow}} \propto \erz{i\uparrow}\erz{i\downarrow}
\ver{i\downarrow}$ and orthonormalize the resulting operators which yields
\begin{subequations}
\label{eq:three_basis}
\begin{alignat}{2}
	\wone{i} &= \sqrt2 \erz{i\uparrow} \\
	\wtwo{i}{j}{k} &= \left(\!\sqrt2\right)^3 \erz{i\uparrow}\left(\erz{j\downarrow}
	\ver{k\downarrow}-\frac{1}{2}\delta_{jk}\right).
\end{alignat}
\end{subequations}
These two operator families comprise $N^3+N$ operators in total. Since the corresponding 
cluster consists of at most three distinct lattice sites we call the ONOB given in 
\Cref{eq:three_basis} the \threeBasis. 
We stress that the \threeBasis{} is invariant under repeated application of 
$\LiouvilleSub{0}{\cdot}$, i.e., the application $\LiouvilleSub{0}{\cdot}$ does not
lead to operators which are not linear combinations of operators in the \threeBasis.
It allows us to reach the same level of description which was reached 
perturbatively by continuous unitary transformations \cite{Moeckel2008,Moeckel2009}.

 We use the \threeBasis{} as a suitable starting point. Next, we modify it to be invariant 
under the application of $\LiouvilleSub{\mathrm{int}}{\cdot}$ because we aim here
at a description of strong interaction quenches where the local
terms dominate. Since the local Hilbert space is four dimensional,
there are at most 15 non-trivial local operators. Due to the symmetries of the
Hamiltonian such as particle number and spin conservation we only need
seven additional operator families with up to nine fermionic creation 
and annihilation operators each, see \Cref{app:basis_operators},
 in order to extend the \threeBasis{}
to the \threePlusBasis{} which is closed under application of $\LiouvilleSub{\mathrm{int}}{\cdot}$.
The \threePlusBasis{} is well suited for strong quenches where hopping can be seen as a pertubation. The basis is exact up to monomials of order $\left(\nicefrac{J}{U}\right)^2$
because it comprises up to three sites and, starting from a single sites, clusters of
three sites require at least two hopping processes.

\section{Observables}
\label{s:observables}

\subsection{Occupation number operator}

The local particle number $n_{i\sigma}(t)$ and its
infinite-time average are the first quantities we address. Of course, particle-conservation
in a translational invariant system tells us that these quantities have to be constant
in time and equal to the filling factor $n$.
 But the advocated approach does not imply this automatically so
that the proper non-dependence on time provides a perfect first test for the accuracy of the approach.
Any deviations of $n_{i\sigma}(t)$ from $n$ can be ascribed to 
approximations induced by the finite bases.
In this way, we can also assess the different performances
 of the \threeBasis{} and the  \threePlusBasis{}.

Due to translational invariance the considered lattice site is arbitrary and will be 
fixed to $i=0$ from now on. The time evolution of the particle number operator is 
thus calculated using
\begin{equation}
	n_{0\uparrow}(t)=\expval{\bzo{0\uparrow}(t)}=\sum_{m,n} h^{(0)*}_m(t) h^{(0)}_n(t) 
	\expval{A_m^\dagger A_n}.
\end{equation}
For the infinite-time average we employ the analytical approach \eqref{eq:n_infty_2}. 

% Seite 49
\begin{figure}[htb]
  {
  \vspace{-.6cm}
  \centering
  \includegraphics[width=1.05\columnwidth,clip]{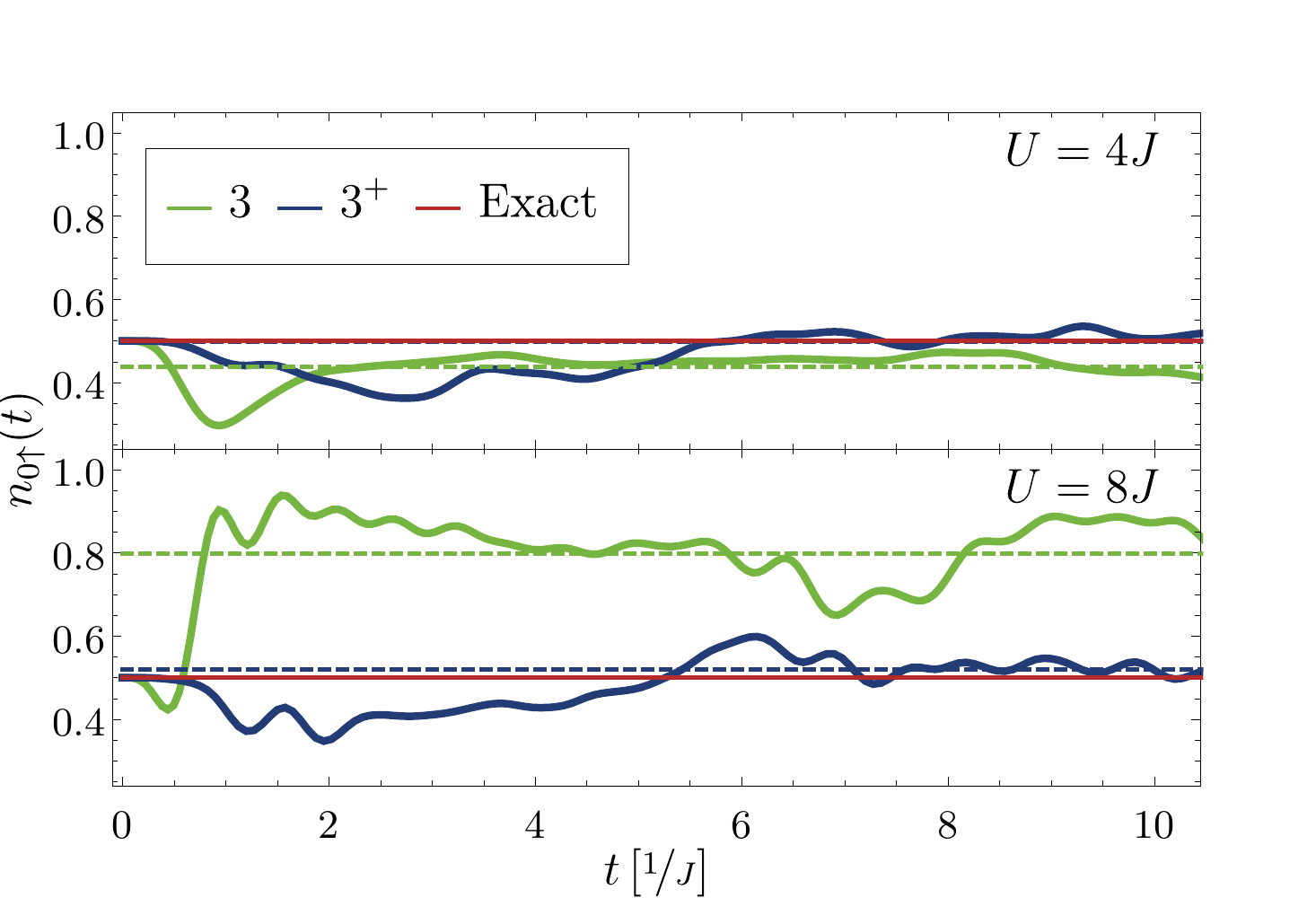}
  \caption{(Color online) Local particle number computed for the \threeBasis{} and the \threePlusBasis{}, for two interaction strengths $U$ at half-filling
	$n=1/2$. The horizontal dashed infinite-time averages $n_\infty$ 
	are depicted for orientation.
  \label{fig:comparison_3_3p_N_12}}}
\end{figure}

\Cref{fig:comparison_3_3p_N_12} compares results from the two ONOBs used. Two qualitatively distinct parameter regimes are studied. For the regime $U=W=4J$ the physical effects of hopping and on-site interaction balance each other which is why the \threePlusBasis{} and the 
\threeBasis{} are expected to lead to qualitatively similar results. 
This agrees with the calculated results. We stress, however,  that the semi-analytically
computed infinite-time average $n_\infty$ of the 
\threePlusBasis{} is noticeably closer to the exact filling $n=1/2$ 
than the equivalent average of the \threeBasis. 

Results for the second parameter regime $U=8J$ 
with a dominating on-site repulsion agree as well with the expectations. Operator 
monomials created by application of $\LiouvilleSub{\text{int}}{\cdot}$ gradually gain 
more weight for $\nicefrac{J}{U}\to0$. Thus, the \threePlusBasis{} is describing the true expectation value of the local particle number significantly better. 
We again emphasize that the inserted infinite-time averages (horizontal dashed lines) 
are not calculated by averaging numerically over $n_{0\uparrow}(t)$ for some time interval.
Instead, they fully rely on the Liouville matrix and the initial condition at $t=0$. 
We checked that $n_\infty$ agrees with the corresponding time averages over $n_{0\uparrow}(t)$ for sufficiently long, but finite times (not shown here).

% Seite 52/53
\begin{figure}[htb]
  {
  \vspace{-1cm}
  \centering
  \includegraphics[width=1.09\columnwidth,clip]{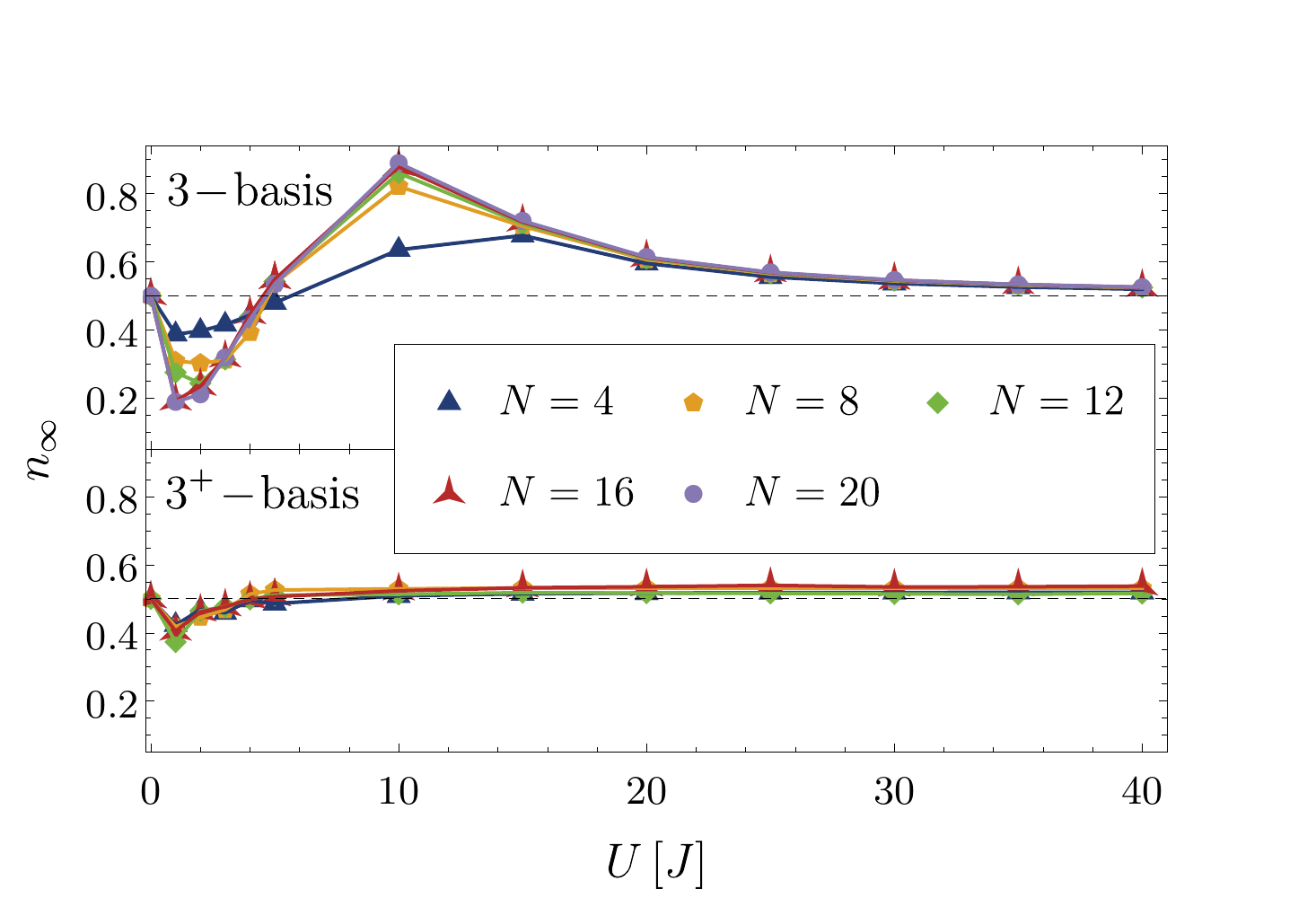}
  \caption{(Color online) Infinite-time averages $n_\infty$ for the \threeBasis{} 
	and \threePlusBasis{} depending on  $U$. The dashed black lines denote the 
	analytically correct result for half-filling. 
	Especially the \threePlusBasis{} is able to keep this value regardless of $U$
	except for a dip around $U\approx J$ while the \threeBasis{} appears not to be
	very reliable. Calculations in the upper (or lower) panel are computed using the 
	Lanczos algorithm for a maximum Krylov space dimension $f$ of a third of the overall Hilbert space
	dimension (or $f\le1000$). A lattice size of $N=20$ was examined for the \threeBasis{} only.
  \label{fig:quantitative_comparison_combined}}}
\end{figure}

To examine the two ONOBs further the infinite-time average $n_\infty$ in 
dependence of $U$ is displayed in \Cref{fig:quantitative_comparison_combined}. 
We highlight that especially in the range of $U\approx 10J$ 
the \threePlusBasis{} is considerably superior to the \threeBasis{} 
regarding the long-term accuracy. 

We conclude that the advocated approach works well. It is systematically controlled
such that the larger operator basis provides better results. For the Hubbard
model under study the \threePlusBasis{} is a very good choice which we will employ
in the remainder of this article for this reason. The high accuracy of the 
directly accessible infinite-time averages, cf.\ 
\Cref{fig:quantitative_comparison_combined}, is very promising, in particular for
relatively strong interaction quenches.

\subsection{Momentum distribution}
\label{ss:momentum_distribution}

The momentum distribution is a key quantity in solid state physics 
and for artificial systems of ultracold atoms in optical lattices.
It can be considered in equilibrium and out-of-equilibrium.
In solid state systems, angular-resolved photoemission spectroscopy
is a tool to measure momentum dependencies \cite{damas03}. But the
determination of the momentum distribution remains difficult because
matrix element effects play an important role and they are not
easy to capture quantitatively.

The  measurement of momentum distributions
in systems of optical lattices is relatively easy. 
At each desired instant of time the optical lattice and any trapping 
potential is switched off suddenly 
so that the particle spread according to their
instant velocities $\vec{v}\propto \vec{k}$. Hence, after a certain
delay time an image of the particle distribution reveals
the momentum distribution just before the sudden release \cite{bloch08}.
Due to its importance we address the time-dependent momentum distribution here.

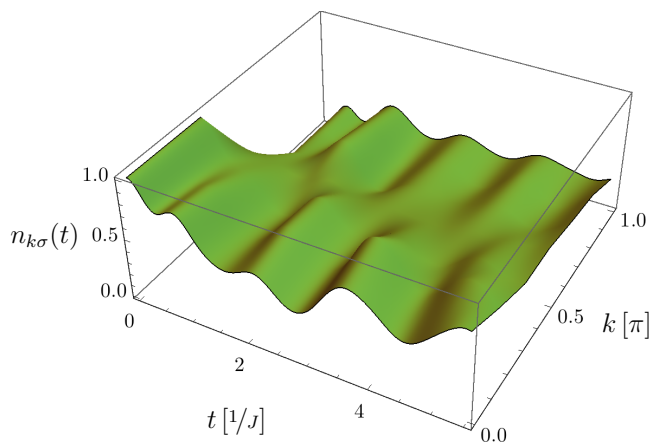
\begin{figure}[htb]
  \hspace{-.7cm}
  \scalebox{.85}{
  \begin{tikzpicture}
  \node (img) {\includegraphics{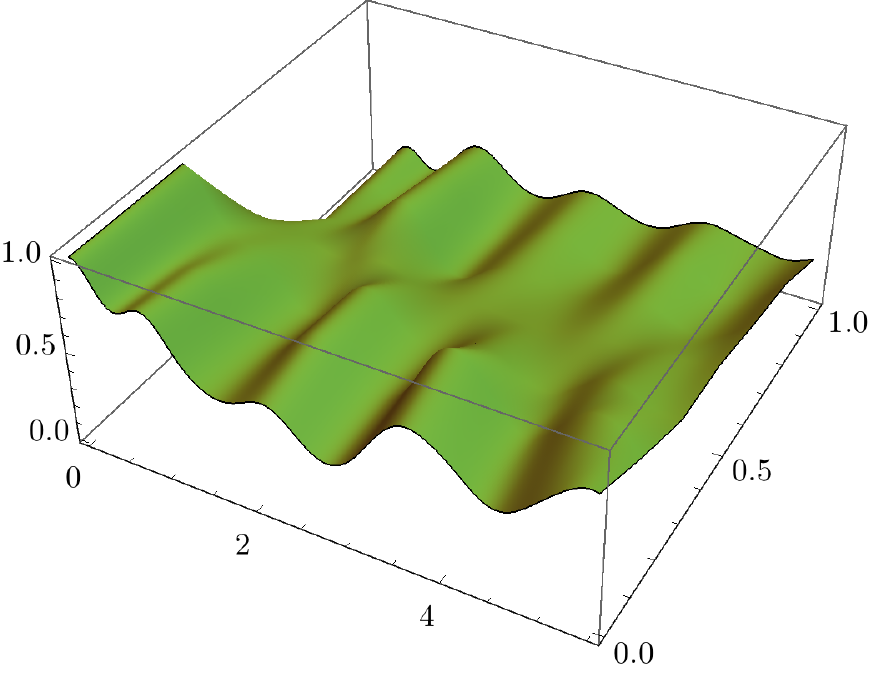}};
  \node[node distance=0cm, xshift=-2cm,yshift=-3cm,font=\color{black}] {\large$t\,[\nicefrac{1}{J}]$};
  \node[node distance=0cm, xshift=4.1cm,yshift=-1.5cm,font=\color{black}] {\large$k\,[\pi]$};
  \node[node distance=0cm, xshift=-5cm,yshift=-.1cm,font=\color{black}] {\large$n_{k\sigma}(t)$};
 \end{tikzpicture}
 } % end scalebox
 \caption{(Color online) Momentum distribution for a lattice of $N=12$ sites after being 
quenched to $U=4J$ vs.\ time. Only the range of $k\in[0,\pi]$ (lattice constant is set to unity)
 is shown due to symmetry. Finite-size features are the washed-out transition between occupied and unoccupied states at $t=0$. Starting from the Fermi sea an out-of-phase oscillation
of the initially occupied and unoccupied states occur with almost featureless 
distributions in between.
  \label{fig:momentum_distribution_U4}}
\end{figure}

\begin{figure}[htb]
\hspace{-.7cm}
\scalebox{.85}{
\begin{tikzpicture}
  \node (img) {\includegraphics{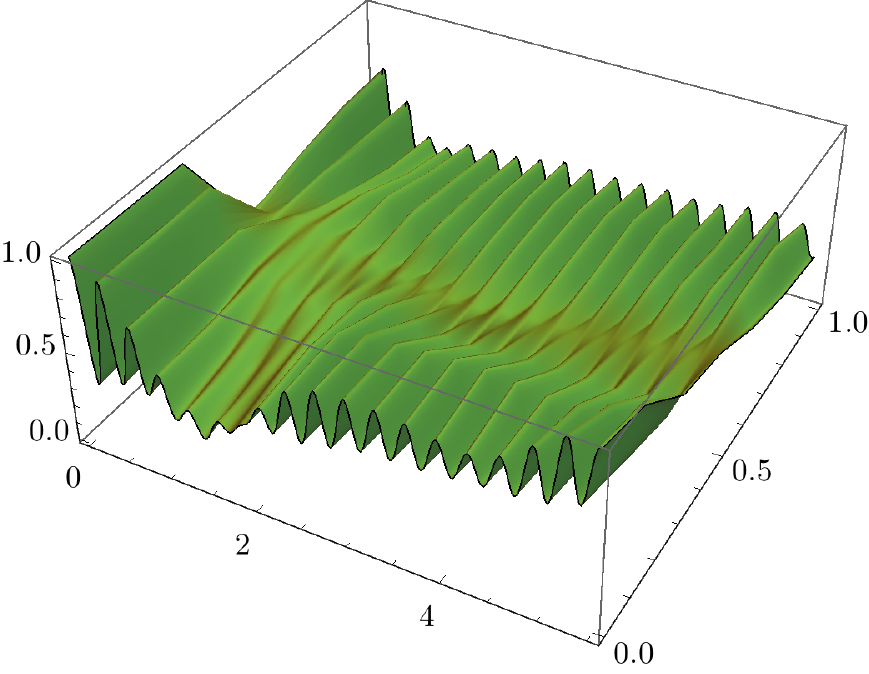}};
  \node[node distance=0cm, xshift=-2cm,yshift=-3cm,font=\color{black}] {\large$t\,[\nicefrac{1}{J}]$};
  \node[node distance=0cm, xshift=4.1cm,yshift=-1.5cm,font=\color{black}] {\large$k\,[\pi]$};
  \node[node distance=0cm, xshift=-5cm,yshift=-.1cm,font=\color{black}] {\large$n_{k\sigma}(t)$};
 \end{tikzpicture}
 } % end scalebox
  \caption{(Color online) Momentum distribution for a lattice of $N=12$ sites after being 
	quenched to $U=20J$ vs.\ time. Qualitatively, the same behavior as in 
	\Cref{fig:momentum_distribution_U4} can be observed. Compared to results of $U=4J$
	the time period of the oscillations is much shorter and there is 
	a huge increase in their amplitudes. At about $t\approx1.7J^{-1} < t_\mathrm{WA}$ 
	a dip in the momentum distribution for excitations of long wavelengths, i.e.,
	$k\approx0$, is visible.
  \label{fig:momentum_distribution_U20}}
\end{figure}
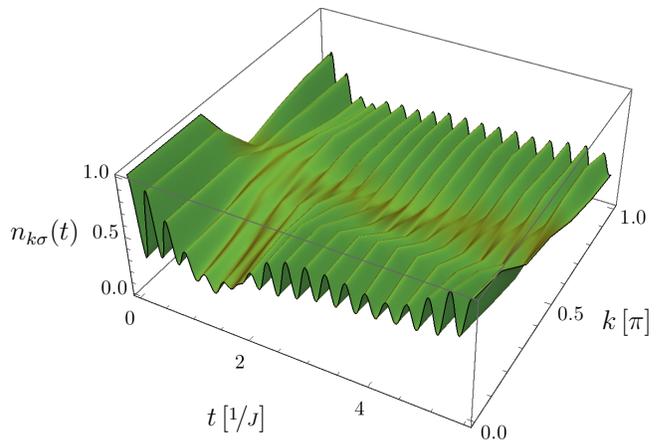

Using the time-dependent non-local correlations
\begin{equation}
	\label{eq:general_exp_corr}
	g_{l\sigma}(t):=\ev{\erz{0\sigma}(t)\ver{l\sigma}(t)}{\text{FS}}
\end{equation}
with their initial values for the Fermi sea for a finite lattice with periodic
boundary conditions of $N$ sites
\begin{equation}
	g_{l\sigma}(t\!=\!0) = \frac{1}{N}\sum_{\mathclap{|k|\leq k_\mathrm{F}}} e^{ikl}
\end{equation}
the time-dependent number of particles with momentum $k$ and 
spin $\sigma$ for time $t$ reads
\begin{equation}
  \label{eq:momentum_distribution_via_corr_functions_compact}
  n_{k\sigma}(t) = \frac{1}{N} \sum_{l} e^{-ikl} g_{l\sigma}(t)
\end{equation}
in the one-dimensional model under consideration. 
The results of the momentum distribution
for the \threePlusBasis{} are shown in 
\Cref{fig:momentum_distribution_U4,fig:momentum_distribution_U20}.

The quench to $U=4J$ is found to be in good agreement with data from an 
iterated equations of motion approach without a unitarity preserving scalar product
\cite{Hamerla2013a,Hamerla2013b} for times up to which the reference results are converged, 
i.e.,  $t\lessapprox 1J^{-1}$. Note that these reference data can be regarded as exact up to
the time threshold up to which they are converged. 
Our data, however, can be trusted up to significantly longer times. 
For the strong quench regime of $U=20J$ there is no data available for comparison. 

There are two main reasons for deviations of our results from the true thermodynamic
behavior. The first one is the use of a truncated basis as discussed in detail above.
In the present work, the truncation is systematically controlled by the choice
of the \threePlusBasis{} which is exact up to and including order $(J/U)^2$.
The second reason is the evaluation of the dynamics on finite chains. 
This implies that wrap-around effects are possible. An operator at site 0 will 
propagate along the chain till it reaches its maximum distance at site $N/2$ after a time 
$t_\mathrm{WA}$. A complete wrap-around occurs after time $t=2t_\mathrm{WA}$.

To assess the reliability of our approach we estimate the maximum speed $v_\mathrm{max}$,
with which information can travel \cite{Lieb1972}, cf.\ \Cref{ss:jump}, by the 
Fermi velocity  $v_\mathrm{F} = 
\left.\dv{k}\epsilon_k\right\vert_{k=k_\mathrm{F}}$ of the 
non-interacting model. We assume that this estimate rather overestimates $v_\mathrm{max}$
because the true velocities in the interacting Hubbard model are generically smaller
due to dressing effects. At half-filling, $v_\mathrm{F}=2J$ for the lattice constant 
set to unity. Hence we arrive at 
\begin{equation}
\label{eq:twa}
t_\mathrm{WA} \approx \frac{N}{4}J^{-1}
\end{equation}
for the time to reach the maximum distance from the initial site on a ring.

Consequently, our results for the momentum distribution in Figs.\ 
\ref{fig:momentum_distribution_U4} and \ref{fig:momentum_distribution_U20}
are taken to be  correct up to at least $t_\mathrm{WA}=3J^{-1}$ 
for the finite lattice size $N=12$ considered. Still, we are able 
to study the momentum distribution in this way for a significantly longer time span 
than previous similar studies \cite{Hamerla2013,Hamerla2013a}.
We will see below, that noticeable wrap-around effects only occur at about
$2t_\mathrm{WA}$ with $t_\mathrm{WA}$ from \eqref{eq:twa}.

The moderate quench to $U=4J=W$ and the strong quench
to $U=20J=5W$ both lead to a considerable redistribution of occupied momenta. 
The initial state is the Fermi sea with clearly occupied and empty momenta.
This distribution is followed directly after the quench
by a rapid increase (decrease) of the average occupation number for $|k|>k_\mathrm{F}$ 
($|k|<k_\mathrm{F}$) with Fermi momentum $k_\mathrm{F}=\nicefrac{\pi}{2}$. 
After this transient rapid change of the occupation of the momenta
 the regions separated by $k_\mathrm{F}$ oscillate 
out-of-phase with an average oscillation period $T$. 
Dedicated analysis, see below, shows that especially in the strong quench regime 
the dominant oscillation possesses a period $T$ of about
\begin{equation}
  \label{eq:rabi_period}
  T=\frac{2\pi}{U}.  
\end{equation}
The physical interpretation is straightforward. In the limit $\nicefrac{J}{U}\to0$ 
the Fermi-Hubbard model is mainly governed by local processes which induce
 Rabi oscillations 
between singly and doubly occupied sites with periods according to \eqref{eq:rabi_period}
because their energy difference is $U$ \cite{Hamerla2013,Hamerla2013a,Hamerla2014}. 
This is analogous to spin precession
about the $z$-axis if initially the spin points along a transversal direction
in the $xy$-plane. In the quenches, the initial state
 is a superposition of the two local possibilities of singly
and doubly occupied sites. Note that at half-filling both, single occupation
and double occupation, are two-fold degenerate. Single occupation due to
the two spin states and double occupation because the completely empty
site acts like a site doubly occupied with holes.

The Rabi oscillations do not die out, but persist on the accessible time scales. 
The momentum distribution for $t<t_\mathrm{WA}$ shows a nearly periodic behavior with little
changes in amplitude and frequencies. This seems to hint at a (quasi-)stationary state 
of a constant momentum distribution being overlaid by oscillations. 
At present, it is not clear whether the oscillations decay for $t\to\infty$
and in which way the amplitude of the oscillations depend
on the system size. Further investigations of the time-dependent momentum distribution
are called for. For the jump at the Fermi surface, we address these issues below.
\begin{figure}[htb]
 {
 \centering
  \includegraphics[width=\columnwidth,clip]{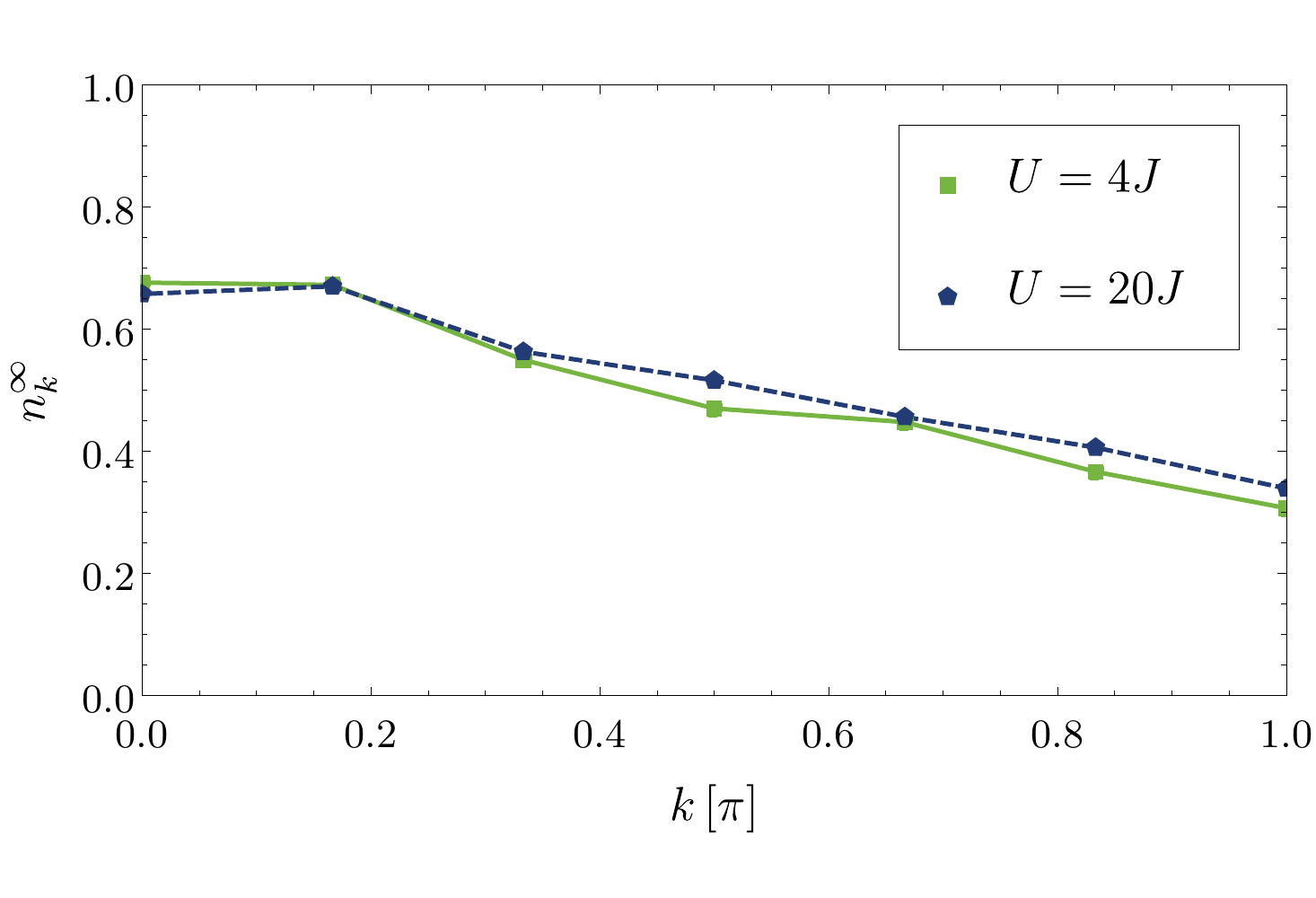}
  \vspace{-.8cm}
  \caption{(Color online) Infinite-time average of the momentum distribution 
	$n_k^\infty$ for two different values of $U$ for a periodic chain of $N=12$ sites. 
	The infinite-time averaged distributions are very similar and featureless. 
	They show comparatively highly occupied momenta for 
	$k>\frac{\pi}{2}$ which were initially completely empty. This calculation is based on a complete diagonalization and does not make use of the Lanczos algorithm.
  \label{fig:mean_momentum_distribution_combined}}
  \vspace{-.5cm}
  }
\end{figure}

To further assess the character of the stationary state in question 
we apply the concept of infinite-time averages to the momentum distribution.
The explicit expression reads
\begin{equation}
  \label{eq:long_term_average_corr_functions}
  g_{l\sigma}^{\infty} = \sum_{\substack{i,j\\\lambda_i=\lambda_j}} \conjugateTransposeSuperscript{\overline{\vec{v}}}{i}{(0)}
	\matrixsymbol{A}\overline{\vec{v}}^{(l)}_j
\end{equation}
for the real space correlations. The momentum distributions ensue by computing the
Fourier series of the $N$ correlation functions \eqref{eq:long_term_average_corr_functions}.
Note that \Cref{eq:long_term_average_corr_functions} is merely a generalization of 
\Cref{eq:n_infty_2}. Consequently, only an adjustment of the initial conditions 
for the annihilation operator is needed. This is what the additional superscripts indicate, cf. \Cref{ss:long_term}.

 Results of this approach are shown in \Cref{fig:mean_momentum_distribution_combined}. 
Only a slight decrease of the average occupation number for increasing momenta is visible. Otherwise the infinite-time momentum distribution is remarkably featureless.
Note that both quenches, the moderate one to $U=4J$ and the strong one to $U=20J$,
result in very much the same momentum distribution.
The results shown are obtained for a finite chain of $12$ sites.
The computation is fairly demanding due to the large dimension of
the operator basis.
Calculation in the thermodynamic limit based on an operator basis truncated by 
a finite range of the operator clusters is an interesting step to be
tackled next.

\subsection{Evolution of the jump at the Fermi surface}
\label{ss:jump}

The striking characteristics of the initial Fermi sea is the discontinuity at
the Fermi wave vector $k_\mathrm{F}$. Hence, we track the temporal evolution of
this jump after the quench quantitatively in order to learn more about
oscillations at finite times, the stationary state for infinite times,
and how the latter is approached.

In a direct calculation on finite lattices the jump is not accessible because
the one-sided limits cannot be computed. Thus we use the elegant detour via the
one-particle contributions $H_n(t)$ yielding the \Cref{eq:jump-time}.

\begin{figure}[htb]
 {
 \vspace{-.9cm}
 \centering
  \includegraphics[width=1.1\columnwidth,clip]{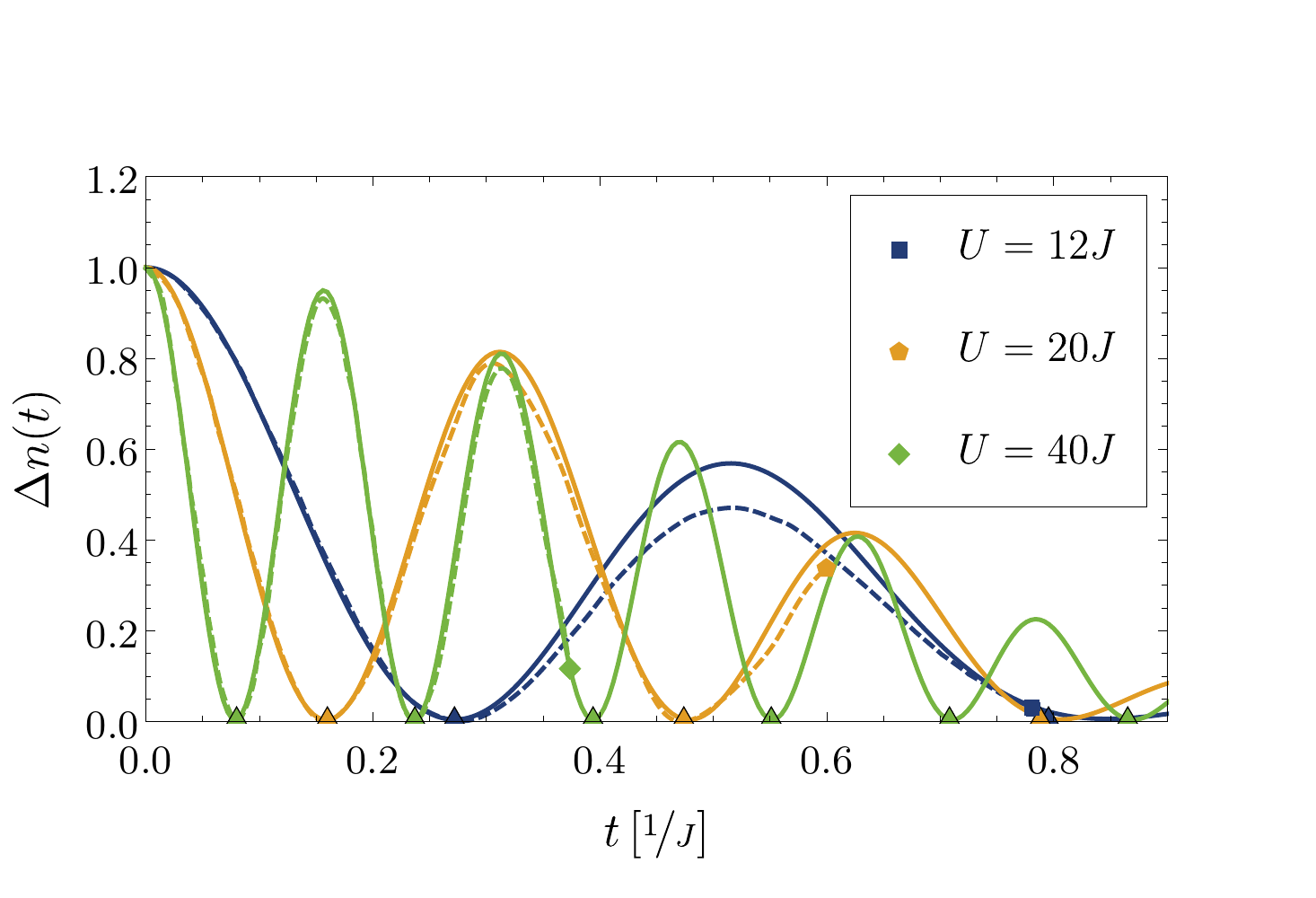}
  \vspace{-.6cm}
  \caption{(Color online) Comparison of results 	obtained using the scalar product for
	operators (solid lines) and results calculated for normal-ordered operators
	without scalar product \cite{Hamerla2013a,Hamerla2013b} (dashed lines). 
	The latter results are calculated in a highly accurate approach, but are only 
	converged up to short times. The respective maximum times are given by the 
	three different plot symbols. Triangles at the $x$-axis mark 
	predictions for zeros of the curve based on the Rabi oscillations 
	\eqref{eq:rabi_period}.
  \label{fig:SH_Delta_strong}}

 }
\end{figure}

For a comparison of results of the iterated equations of motion approach 
using the scalar product given in \Cref{eq:frobenius_scalar_product} 
and not using it \cite{Hamerla2013,Hamerla2013a,Hamerla2013b,Hamerla2014} we display
\Cref{fig:SH_Delta_strong}. The data depicted by the dashed curves
 is highly accurate where it is
 converged, i.e., for short time spans only. The time range shown and 
used for comparison is adapted accordingly.

Two striking phenomena are to be observed. First, for all interaction strengths 
the zeros of each curve agree for both methods extremely well. 
In the considered regime of quenches to strong interactions we see that the
simple estimate for the period of the observed Rabi oscillations \eqref{eq:rabi_period}
matches the zeros very well, see the triangular symbols at the bottom axis. The
first symbol is put at the first zero of the computed curves; the following
symbols are placed at multiples of the Rabi period $T$. Prediction and actual data
agree remarkably well. 

Second, the amplitudes for both approaches nearly coincide for 
$U\ge20 J$. This shows that the scalar product method 
yields accurate results in the limit of strong quenches. 
This provides two advantages: (i) the operator basis to be considered, though large,
is much smaller in the present approach than in the previous approach.
(ii) The present approach is able to capture noticeably longer time spans 
so that it renders more extensive examinations possible, for instance
addressing relaxation behavior.

% Finite size effects mit Verifikation von t_WA
\begin{figure}[htb]
 {
 \vspace{-.6cm}
 \centering
  \includegraphics[width=1.1\columnwidth,clip]{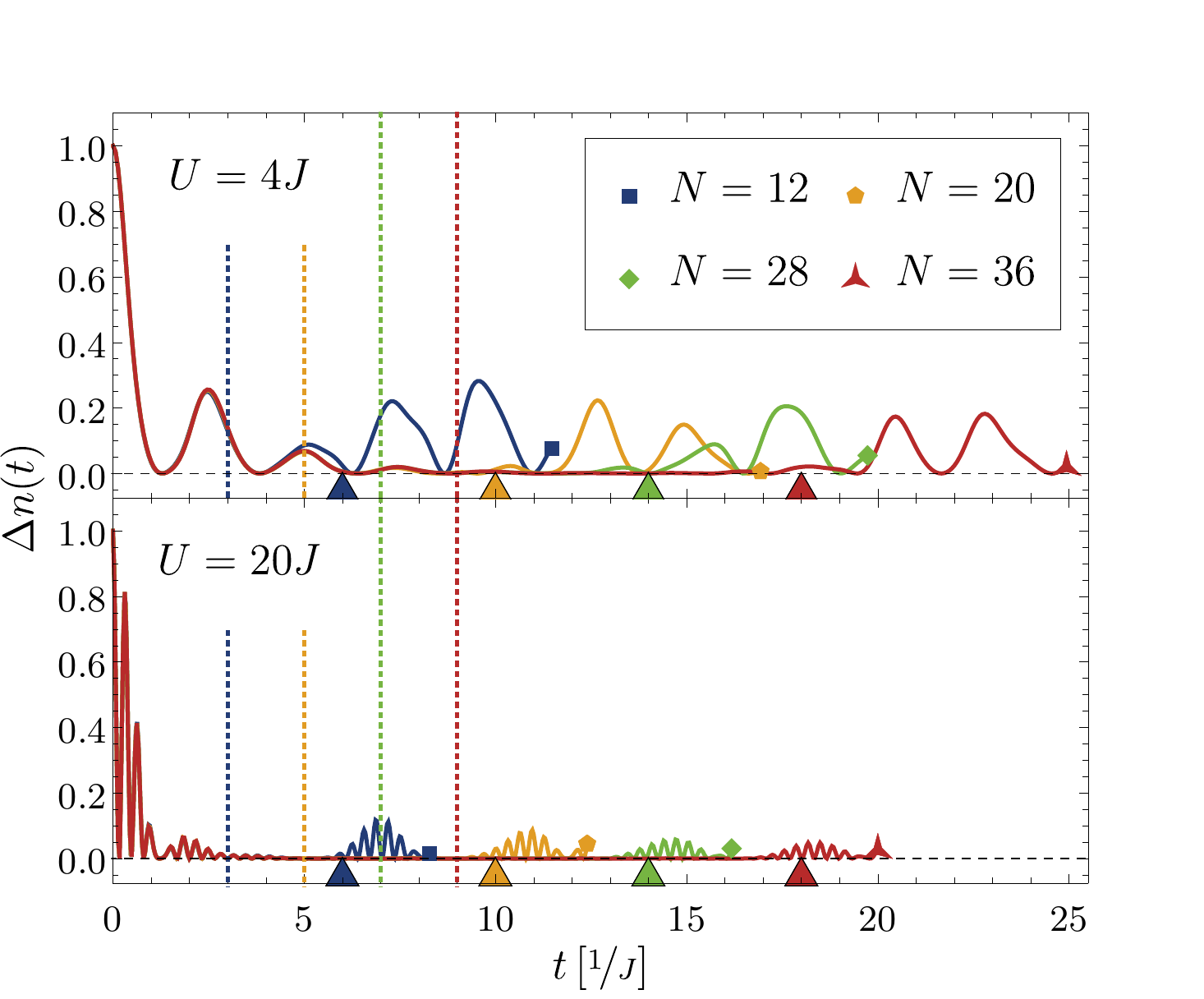}
  \caption{(Color online) Jump at the Fermi surface for two values of $U$ (upper/lower panel)
	in dependence on the lattice size $N$. Finite-size revivals occur the earlier the smaller 
	the lattice is. Once a finite-size revival has occurred the data is cropped by hand. The
	corresponding end-of-curve is denoted by the four different glyphs. 
	Dashed vertical lines mark $t_\mathrm{WA}$, i.e., the time up to which results are 
	estimated to be independent of $N$, colored triangles mark the instants $2t_\mathrm{WA}$
	which indeed almost coincide with the onset of the spurious finite-size revivals.
  \label{fig:delta_finite_size}}}
\end{figure}

For an in-depth analysis of the impact of finite-size effects on the results 
we calculate the jumps at the Fermi surface for different lattice sizes $N$.
The results are shown in \Cref{fig:delta_finite_size}. The initial evolutions of the jump  are identical for all sizes. The jump starts at its maximum value of unity 
and decreases very briefly thereafter in an oscillatory manner.
We see a collapse-and-revival  phenomenon as has been observed before\cite{Hamerla2014} 
until the jump vanishes completely.

Upon increasing $N$ the finite-size revivals occur later and later. The time instant at which
these spurious features set in is estimated very well by $2t_\mathrm{WA}$,
see triangles at the bottom of both panels in \Cref{fig:delta_finite_size}.
For sufficiently large $N$, the jump has essentially vanished before any
finite-size effect sets in.

Interestingly, the evolution up to $2t_\mathrm{WA}$ is almost independent
of the system size. This observation is very promising because it implies that
we are able to capture the essential dynamics of the Fermi jump by
considering finite systems. This allows for a reduction of numerical effort 
in future studies for cases where the behavior up to specific times is needed only.

We conclude from \Cref{fig:delta_finite_size}
that the jump of the Fermi surfaces vanishes very quickly. 
One is tempted to conclude that the system relaxes on these short time scales.
But inspection of \Cref{fig:momentum_distribution_U4,fig:momentum_distribution_U20}
reveals that significant oscillations still take place at momenta
far away from the Fermi wave vector $k_\mathrm{F}$ when the Fermi jump 
has already disappeared.
Hence, further studies of these oscillations and their dependence on
time, interaction, and system size is necessary, but left for future research.

Investigating \Cref{fig:delta_finite_size} we see that the decaying oscillations
are governed by more than one single frequency. In particular in the lower panel
displaying data for a quench to $U=20J$ one discerns a beating in the oscillations
with a zero amplitude at about $t\approx 1.2/J$. This is a clear indication 
of the presence of at least two frequencies. To analyze the relevant frequencies
in the evolution of the Fermi jump systematically we use Fourier analysis.
But it is hampered by the discontinuous onset at $t=0$ where the signal starts.
Since we want to focus only on the frequency content we transform
the symmetrized signal $\Delta n(t)+ \Delta n(-t)$, which we additionally smoothen by means of a low-pass Gaussian filter, by fast Fourier transform. 

\begin{figure}[htb]
 {
 \vspace{-.9cm}
 \centering
  \includegraphics[width=1.06\columnwidth,clip]{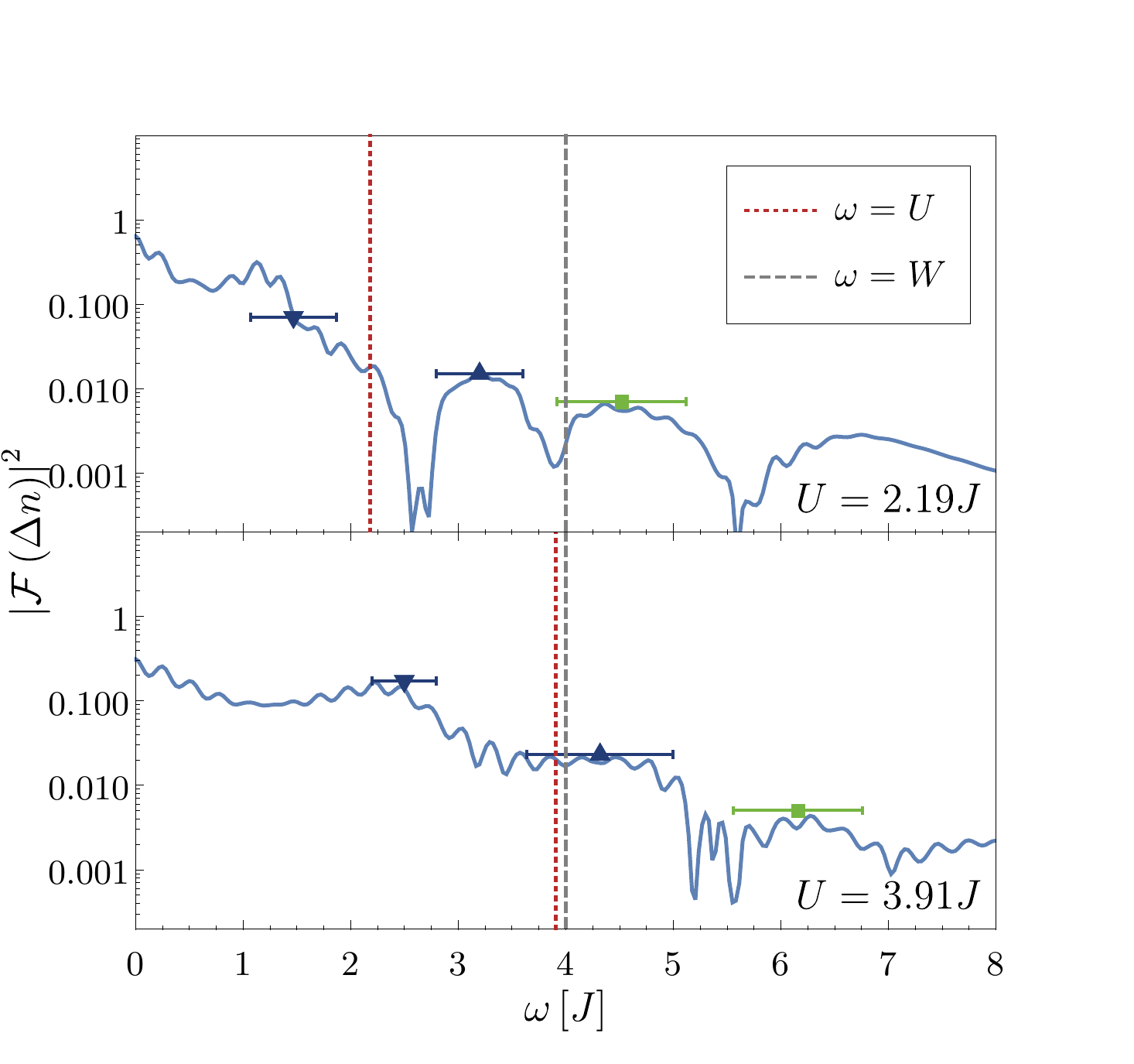}
  \vspace{-.5cm}
  \caption{(Color online) 
	Spectra as obtained by Fourier transform from the time-dependent
	symmetrized Fermi jump for weaker interaction quenches. The symbols with
	error bars indicate spectral features which we read off.
	  \label{fig:spectra1}}}
\end{figure}

\begin{figure}[htb]
 {
 \vspace{-.8cm}
 \centering
  {\includegraphics[width=1.06\columnwidth,clip]{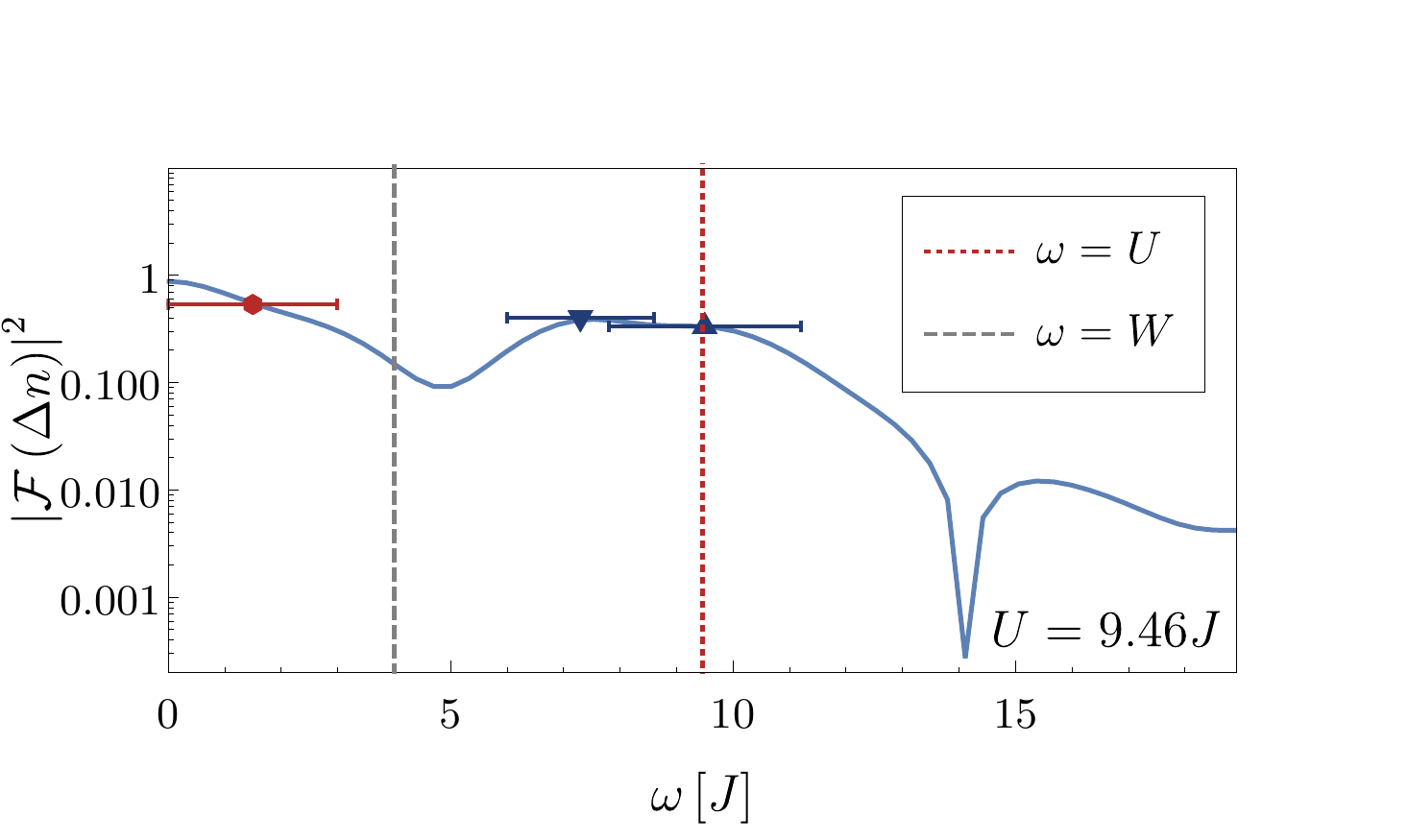}\\
  \vspace{-.7cm}
	\includegraphics[width=1.06\columnwidth,clip]{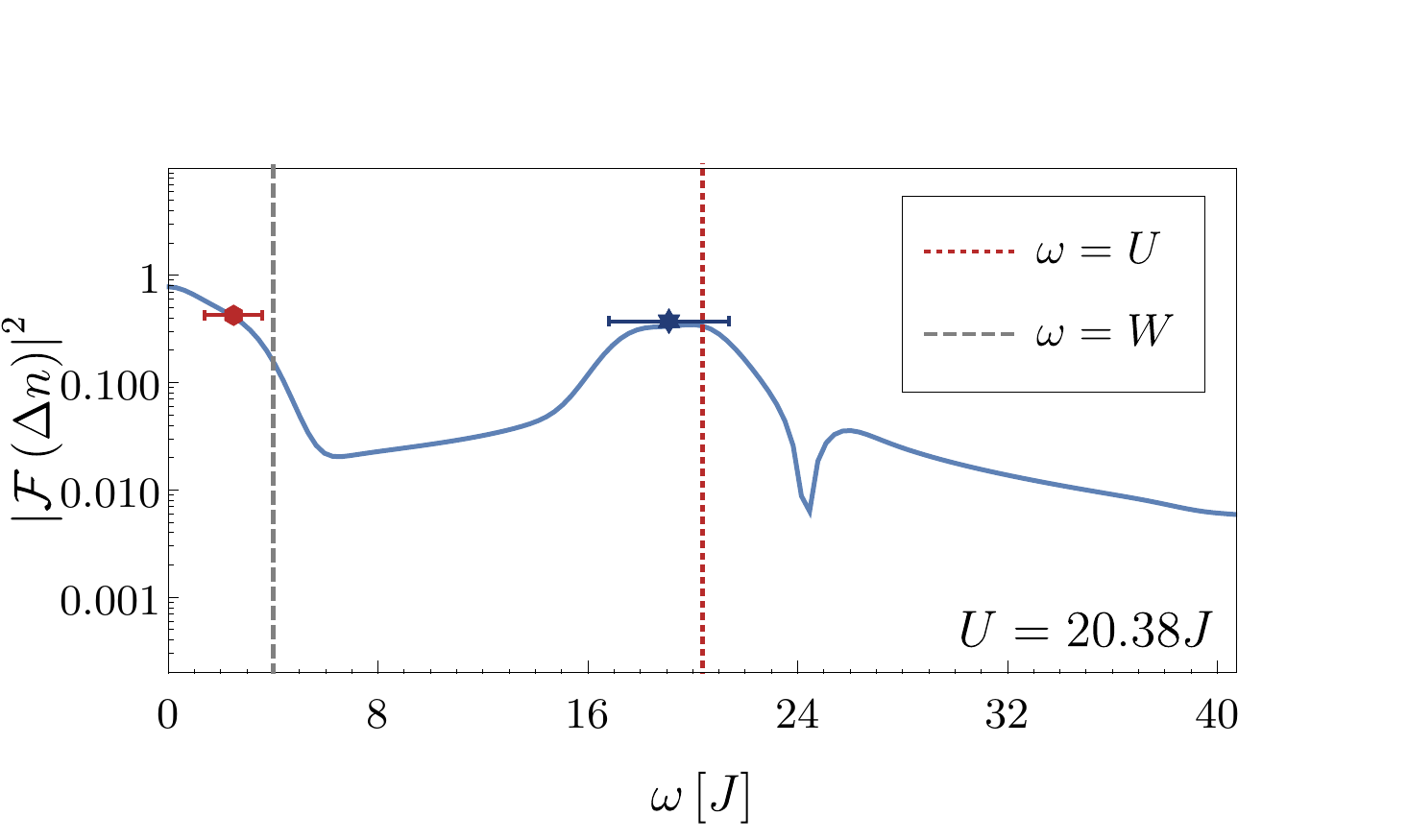}
	}
	\vspace{-.5cm}
  \caption{(Color online) 
	Spectra as obtained by Fourier transform from the time-dependent
	symmetrized Fermi jump for stronger interaction quenches. The symbols with
	error bars indicate spectral features which we read off.
	  \label{fig:spectra2}}
	  \vspace{-.3cm}}
\end{figure}

\Cref{fig:spectra1,fig:spectra2} display the squares of the absolute values of the resulting 
spectra for various quenches in logarithmic plots. The value of the quenched interaction
is indicated in the panels. The colored symbols with error bars show the frequencies
which we read off. Mostly they indicated peaks, but also shoulders,
see the low-frequency feature in the upper panel of \Cref{fig:spectra1}. 
We choose to read off this
feature because at even smaller interaction only the shoulder can be identified.

The vertical dashed lines show two typical energies of the system, namely the
band width $W$ and the interaction strength $U$ for orientation and comparison.
It appears that both of them show up in the spectral features, i.e., 
spectral features occur at frequencies which are close to $W$ or $U$.

Note that different symbols are chosen to display different spectral structures.
Upon increasing $U$, the high-frequency feature shown using a green square shifts to higher and higher values,
but becomes less and less significant, see also \Cref{fig:spectra2}. 
Beyond a certain value of $U$ it
does not appear anymore. In parallel, a low-frequency feature appears which 
was not discernible before. We denote it by the red hexagons in \Cref{fig:spectra2}.
Comparing the upper and the lower panel of \Cref{fig:spectra2} it appears
that the two peaks indicated by blue triangles, which are still discernible
in the upper panel, are merged in the lower panel. So they are denoted by a  single
symbol formed from both triangles.

\begin{figure}[htb]
{\centering
  \includegraphics[width=1.02\columnwidth,clip]{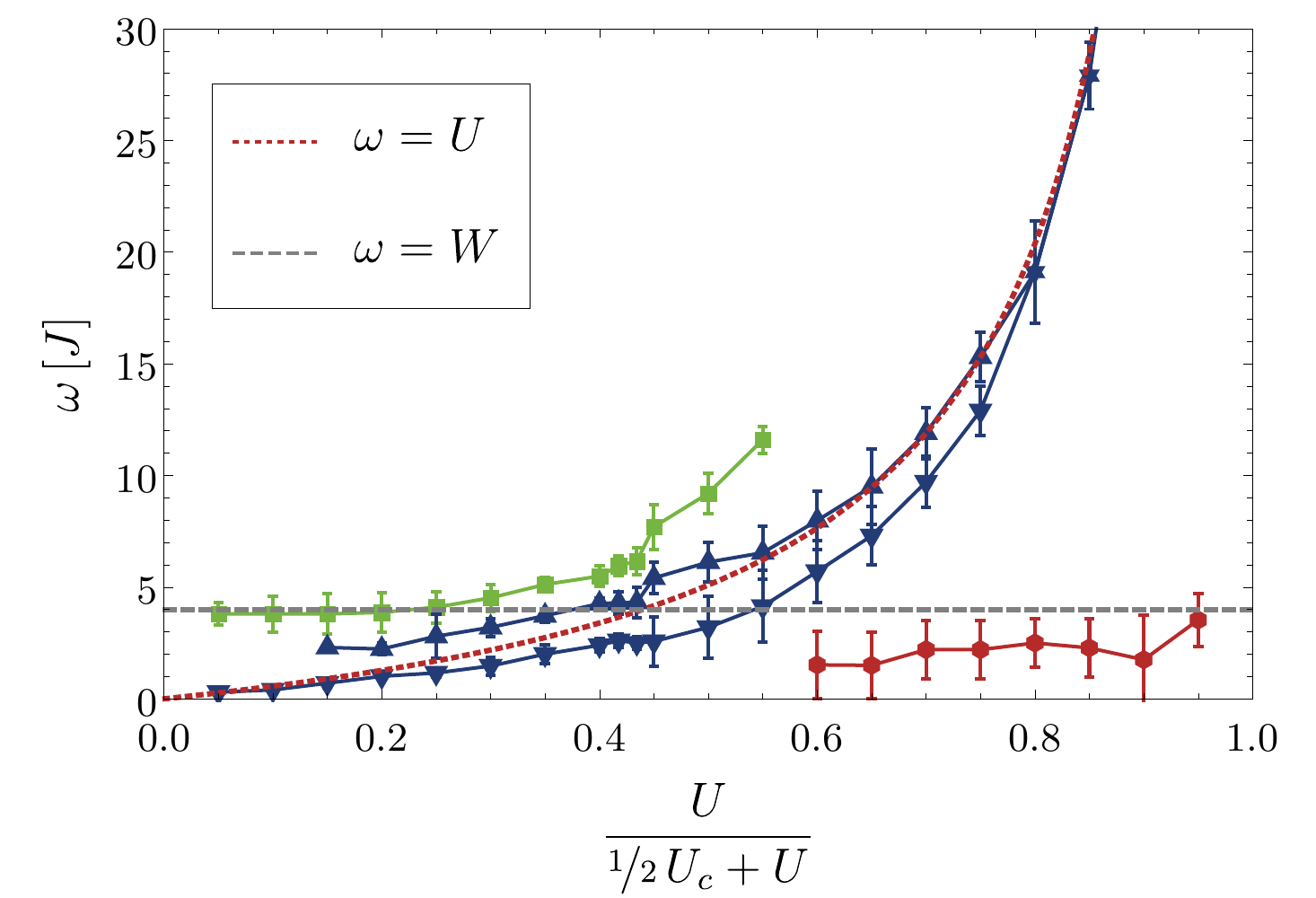}
  \caption{(Color online) Frequencies of the most important spectral features
	in the spectra of $\Delta n(t)$ plotted as functions of the interaction strength.
	The latter is given in a compactified form with $U_c=-8E_\mathrm{kin}/N$
	where $E_\mathrm{kin}/N=-W/\pi$ is the kinetic energy per site of the 
	half-filled non-interacting system. This allows us to show 
		the full range from $U=0$ to $U=\infty$.
	The error bars are determined approximately  by the half-widths at half maximum of the peaks. The frequency of local Rabi oscillations, i.e., $\omega=U$, and the band width, i.e., 
	$\omega=W$, are depicted for comparison as dashed lines.
  \label{fig:dynamical_transition}}
  \vspace{-.3cm}}
\end{figure}

We analyzed many more spectra than the four shown here explicitly. The data is compiled
in  \Cref{fig:dynamical_transition}. In total, we identified four relevant spectral 
features. At weak quenches there is clearly one feature given by the band width $W$,
shown by the green curve in \Cref{fig:dynamical_transition}. Besides this feature,
there are two features located at frequencies  above and below the local Rabi
frequency $U$ (blue curves). The lower frequency almost coincides with $U$ for
weak quenches.  We stress that the features for weak quenches must
be regarded with some caution because the choice of the ONOB is designed for
strong quenches.

For strong quenches, there occurs a low-frequency feature depicted by the red
curve. It does not coincide quantitatively with the band width $W$, but
it is close to it within a factor of two. Given the difficulty to extract the proper
frequency for the low-frequency feature, see \Cref{fig:spectra2}, the quantitative
deviations are not surprising. The high-frequency feature clearly matches the local
Rabi frequency $U$ almost quantitatively. The two spectral feature above and below
$U$ merge for larger $U$, at least they can no longer be detected separately,
see lower panel of \Cref{fig:spectra2}.

The intermediate parameter region $U\approx U_c/2$ in
\Cref{fig:dynamical_transition} is of particular interest. In previous analyses 
\cite{Eckstein2009,Schiro2010,Schiro2011,Hamerla2013} it seemed as if there
were two qualitatively distinct regimes for weak and for strong quenches,
separated by a dynamical phase transition. The results in 
\Cref{fig:dynamical_transition} question this interpretation.
We recall that the previous pieces of evidence were justified for
infinite dimensions \cite{Eckstein2009}, restricted
in the accessible times \cite{Hamerla2013}, or they
were based on a variational ansatz neglecting a significant part of 
quantum fluctuations \cite{Schiro2010,Schiro2011}.

\Cref{fig:dynamical_transition} points towards a crossover. Indeed,
different spectral features dominate for weak and strong quenches.
But there is no sharp, singular transition between these two regimes.
Instead, the weight of the different spectral features shifts so that
$W$ is more relevant for weak quenches while $U$ dominates for strong quenches.
This issue as well deserves further investigation.
The situation in higher dimensions, for instance in $d=2$, would be
particularly interesting.

\subsection{Infinite-time average of the Fermi jump}

In the previous subsection, we discussed the temporal evolution
of the jump at the Fermi level. Here we finish our analysis of the Fermi
jump by presenting its infinite-time averages. The necessary formula
has been given in \Cref{eq:Delta_n_infty}. 

% Delta n_infty und erzählen, dass das in Übereinstimmung mit der -> 0 Vermutung ist.
\begin{figure}[htb]
{

\centering
  \includegraphics[width=\columnwidth,clip]{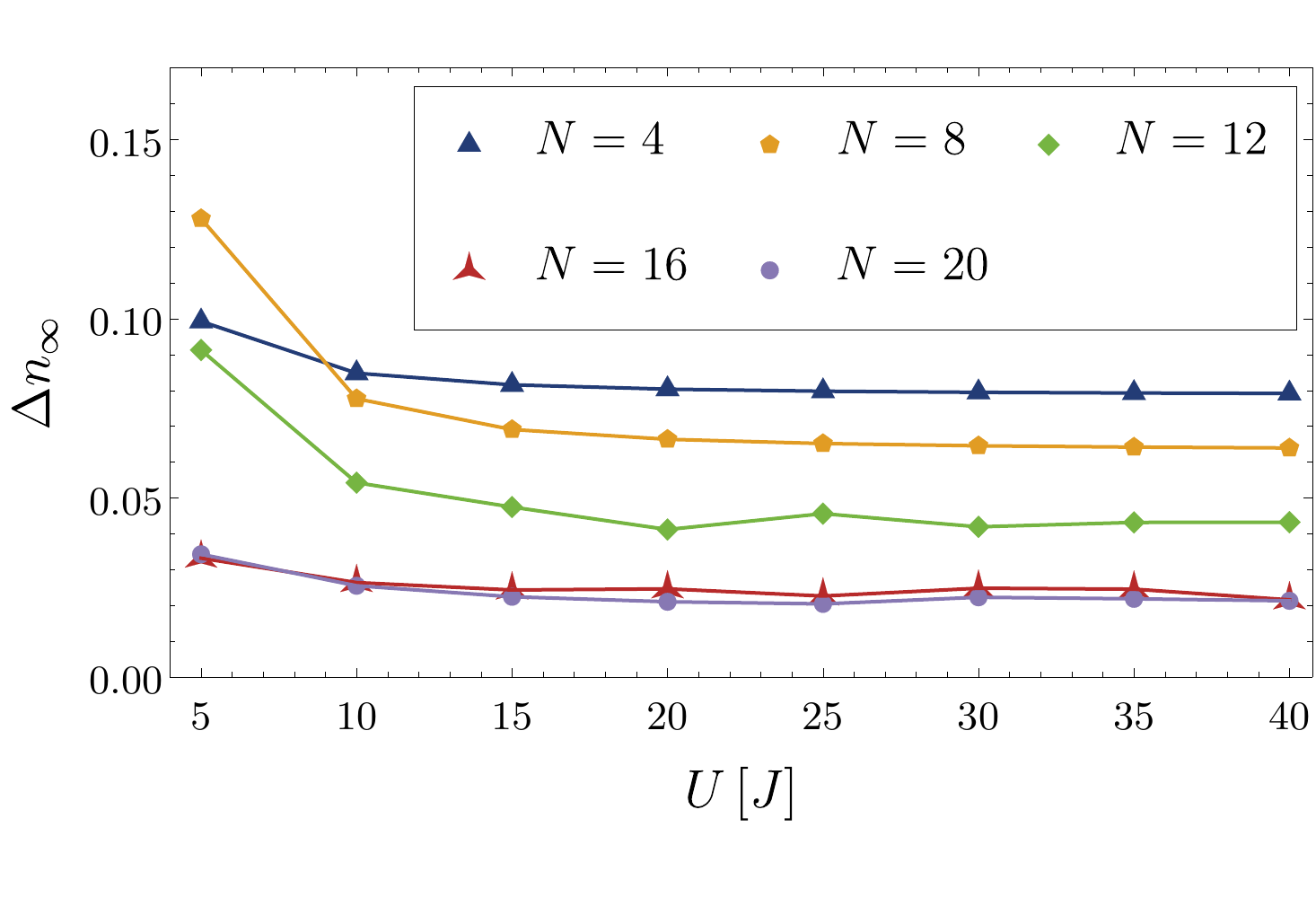}
  \vspace{-.8cm}
   \caption{(Color online) Infinite-time averages of the jump $\Delta n_\infty$ 
	calculated from \eqref{eq:Delta_n_infty}. 
	No pronounced dependence on $U$ is found except for weaker quenches $U<2W=8J$,
	see discussion in main text. The calculation shown is performed using the Lanczos algorithm with a maximum Krylov space dimension $f\le1000$.
  \label{fig:delta_infty_fs}}
  \vspace{-.3cm}}
\end{figure}

The ensuing data for various
system sizes is displayed in \Cref{fig:delta_infty_fs} in dependence on 
the quenched interaction strength $U$. Clearly, for larger
values of $U$, no noticeable dependence on $U$ arises.
The stronger dependence for smaller values of $U\approx5J$ can be attributed
to the slower decay of the jump, see \Cref{fig:delta_finite_size}. If the decay is slow,
the revivals are larger. Hence, for smaller system sizes the jump seems to be
larger. We emphasize that the infinite-time averages also comprise all the
effects of revivals. Consequently, it is explainable that the infinite-time averages
are finite in spite of the observation that they vanish rapidly,
see \Cref{fig:delta_finite_size}.

\begin{figure}[htb]
{
\vspace{-.4cm}
\centering
  \includegraphics[width=\columnwidth,clip]{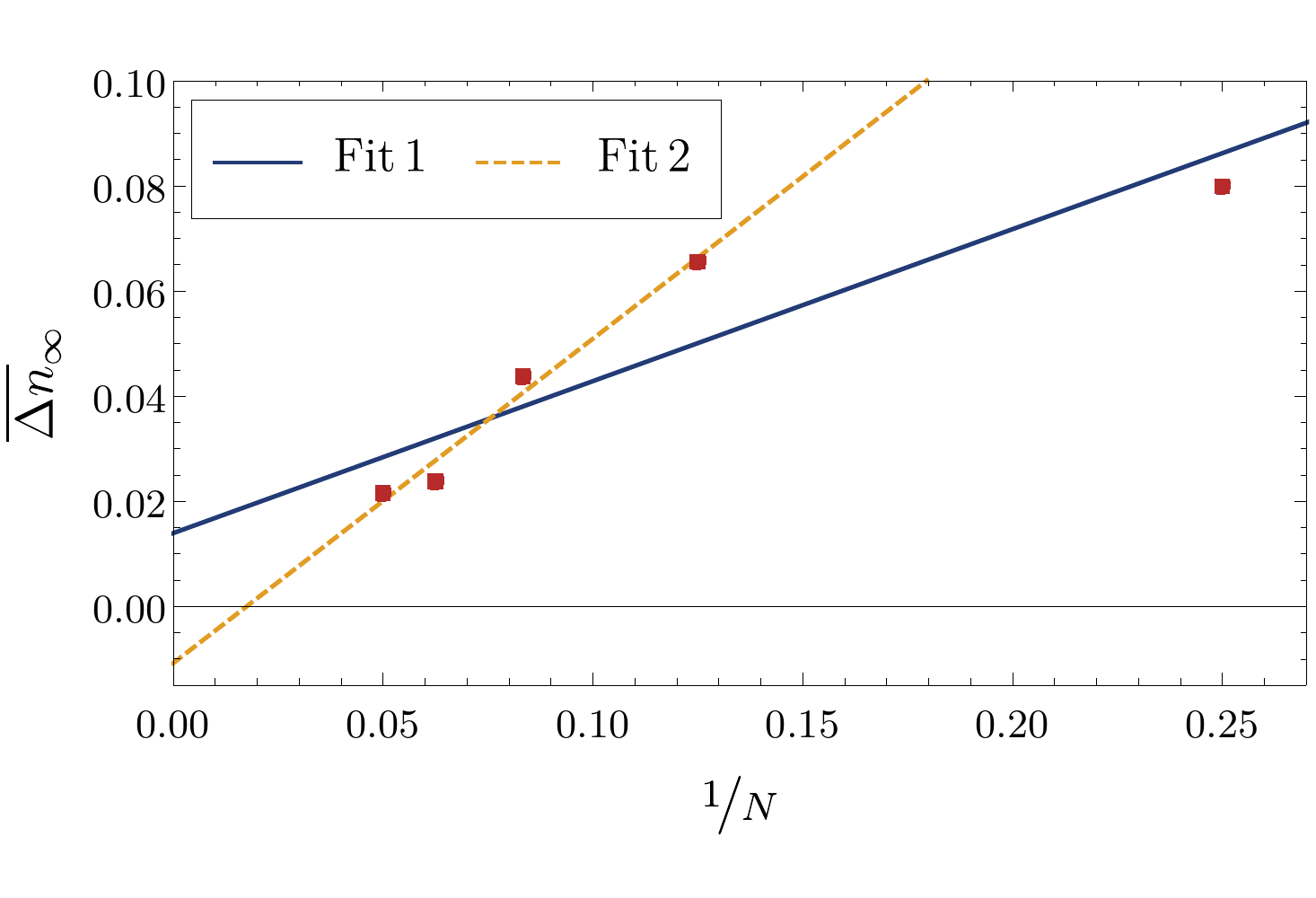}
  \vspace{-.6cm}
  \caption{(Color online) Mean values $\overline{\Delta n_\infty}$ of the infinite-time jump $\Delta n_\infty$
	averaged between $U=15J$ and	$U=40J$ vs.\ the inverse lattice size. The linear fit 
	1 (2) includes (excludes) the data point for $N=4$.
  \label{fig:delta_infty_extrapolation}}
  \vspace{-.3cm}}
\end{figure}

If the above sketched view is correct, larger systems with later and weaker
revivals should show smaller values of $\Delta n_\infty$. Thus, we aim
at a finite-size extrapolation. In order not to do such an extrapolation 
for many different values of $U$ we use the fact that they hardly
depend on $U$ as long as it is large, see \Cref{fig:delta_infty_fs}.
We average $\Delta n_\infty$ in the interaction interval $U\in[15J,40J]$ and 
plot the results as function of the inverse system size $1/N$ in 
\Cref{fig:delta_infty_extrapolation}.

For extrapolation we employ linear fits. There are not enough points
and they scatter a bit so that the accuracy of the extrapolation is limited
to about 2\%.
But fit 2, excluding the very small system size of $N=4$ sites, yields
a value close to zero, though negative, but very small. The fit 1, including
the case $N=4$, yields a small positive value. The difference of both extrapolated
values yields an estimate for the accuracy which is about $0.02$. Clearly, the
extrapolations are fully consistent with $\Delta n_\infty=0$ for the
thermodynamic limit, i.e., for $N=\infty$. We stress that this finding is
the same that we concluded already previously from the analysis of 
\Cref{fig:delta_finite_size}.

\section{Summary}
\label{s:summary}

In the present article, progress has been achieved in two main domains.

The first domain is conceptual. We presented a general approach to compute
time-dependent expectation values based on time-dependent operators
in the Heisenberg picture with truncated bases. 
The key idea is to choose an appropriate
scalar product for operators such that the Liouville operator, i.e., 
commutation with the Hamiltonian, is self-adjoint. We presented such
a suitable choice based on the  Frobenius scalar product for a fermionic model.
An application to spin models has been published elsewhere \cite{Rohrig2017}. 
As noted before \cite{Kalthoff2017}, however,
this approach only works as such for finite local Hilbert spaces, i.e., for lattice
models of fermions and spins.

On the conceptual side, we, moreover, showed that the self-adjoint Liouville operator
implies a unitarity of the time evolution on the \emph{operator} level.
Generically, this implies oscillatory behavior as it has to be in quantum
mechanics. We discussed comprehensively that the usual unitarity
on the level of \emph{states} is clearly distinct. Indeed, the latter implies
operator unitarity, but the operator unitarity does not imply state unitarity.

Based on the operator unitary time evolution  we used the concept
of stationary phases to derive equations which directly express 
the expectation values of the (quasi-)stationary state to which the
system converges for infinite time. The prefix ``quasi'' expresses that
one cannot be sure that this state does not still show oscillations.
In finite systems, we would indeed expect that oscillations persist.
But they typically decrease upon increasing system size.

The second domain is the concrete application of the 
abstract concepts developed in the first domain to
a fermionic model. For simplicity, we choose the one-dimensional
Hubbard model and studied an interaction quench from the non-interacting
Fermi sea to some finite interaction $U$. The integrability of the
model is not used at any stage.

We conceived the \threePlusBasis{}. This basis describes all possible physical 
processes which can occur on up to three lattice sites. It is designed to capture 
the limit of strong interactions and small hopping; it is exact up to order $(J/U)^2$.
The performance of this orthonormal operator basis (ONOB) was tested by
computing the local particle number. Rigorously, it should be constant
and equal to the filling factor.
In the truncated approach, however, this need not be the case.
But we showed that the \threePlusBasis{} reproduces the filling factor
to good accuracy. For large values of $U$ the agreement is even
very good.

Then, we computed the time evolution of the momentum distribution. 
It is dominated by out-of-phase oscillations of the initially occupied and 
unoccupied momenta. These oscillations correspond to Rabi oscillations,
well-known from two-level systems. They are reminiscent of the collapse-and-revival
scenario in bosonic systems \cite{Greiner2002} and they were observed
before in the one-dimensional \cite{Hamerla2013}, the  two-dimensional Hubbard model
\cite{Hamerla2014}, and the infinite-dimensional Hubbard model \cite{Eckstein2009}.

Finally, we focussed on the Fermi jump, i.e., the discontinuity
of the momentum distribution at the Fermi wave vectors. Here we showed
that the oscillations quickly die out. Finite-size effects are
completely controllable. A systematic frequency analysis shows
that two important frequencies/energies dominate: the band width $W$
and the interaction $U$. Interestingly, we do not find a singular
dynamic phase transition as function of the quenched interaction,
but a smooth crossover. Spectral features gain and lose weight,
but they do not pop up or vanish suddenly. This is in contrast
to previous interpretations. This progress has become possible due to the 
significantly longer accessible times thanks to the conceptual
progress in the first domain.

\section{Outlook}
\label{s:outlook}

Optimizing the code will enable one to tackle significantly larger systems.
Then, a quantitative analysis of the decay times at the Fermi wave vector
and far away from it will be possible and is called for.
This can be done in one dimension, but since no special feature
of the one-dimensionality has been exploited, the same objectives
can be pursued for two-dimensional systems. Of course, the accessible
linear sizes will be more limited.

Another important route to follow is to enlarge the operator basis,
for instance from three to five sites. Then, it should be possible
to study the relaxation of weak quenches, i.e., the regime $U/J\ll 1$.
This regime is in principle also accessible to perturbative, diagrammatic
approaches which opens the way to direct comparisons.

Finally, we stress that different initial conditions can also be implemented
by adjusting the matrix $\matrixsymbol{A}$. From the conceptual point of view,
this is straightforward. For instance, initial states which are
relatively simple product states can be accounted for easily, 
even if they break a symmetry, e.g., spin or translational symmetry.
More work has to be done in order to determine the initial correlations
in strongly correlated systems. These require an equilibrium calculation
in the first place in order to know the initial conditions.

In summary, there is a plethora of questions to address so that
we are confident that the field of non-equilibrium quantum physics
will continue to thrive.

\begin{acknowledgments} 
We gratefully acknowledge financial support of the DFG in project space UH 90-13/1. 
Furthermore, we thank J\"org B\"unemann, Joachim Stolze, and Fabian K\"{o}hler
 for helpful discussions and the latter one also for provision of data for comparison in the early stages of the project as well as for technical advice.
\end{acknowledgments}

\begin{widetext}
\begin{appendix}
\section{Basis operators}
\label{app:basis_operators}

For completeness, we provide the full \threePlusBasis{} which is used mainly
in this article. This basis is closed, i.e., invariant under iterated 
application of $\LiouvilleSub{\text{int}}{\cdot}$. Consequently, it is well-suited 
to study the regime $\nicefrac{J}{U}\ll1$, i.e., comparatively strong quenches. 

The following orthonormal basis operators have to be added to the two given in 
 \Cref{eq:three_basis}, eventually leading to nine different operator families 
in total in the complete  \threePlusBasis{}.

Restrictions regarding the site indices apply. The following operators exist 
for three distinct indices $i\neq j\neq k$, $i\neq k$. 
Note that the operators $\wthree{i}{j}{k}$ also exist for the case where only $i$ is distinct from the other two indices. The same applies to $\wfour{i}{j}{k}$ for the index $j$ and $\wfive{i}{j}{k}$ for the index $k$. 

\begin{subequations}
\begin{alignat}{7}
	\wthree{i}{j}{k} &= \left(\!\sqrt2\right)^5 \erz{i\uparrow}\left(\bzoMod{i\downarrow}
	-\frac{1}{2}\right)\left(\erz{j\downarrow}\ver{k\downarrow}-\frac{1}{2}
	\delta_{jk}\right) \label{eq:threePlus_basis_wthree}
	\\
	\wfour{i}{j}{k} &= \left(\!\sqrt2\right)^5 \erz{i\uparrow}\erz{j\downarrow}
	\left(\bzoMod{j\uparrow}-\frac{1}{2}\right)\ver{k\downarrow} 
	\\
	\wfive{i}{j}{k} &= \left(\!\sqrt2\right)^5 \erz{i\uparrow}\erz{j\downarrow}
	\ver{k\downarrow}\left(\bzoMod{k\uparrow}-\frac{1}{2}\right) 
	\\
	\wsix{i}{j}{k} &= \left(\!\sqrt2\right)^7 \erz{i\uparrow}\left(\bzoMod{i\downarrow}
	-\frac{1}{2}\right)\erz{j\downarrow}\left(\bzoMod{j\uparrow}-\frac{1}{2}\right)
	\ver{k\downarrow} 
	\\
	\wseven{i}{j}{k} &= \left(\!\sqrt2\right)^7 \erz{i\uparrow}
	\left(\bzoMod{i\downarrow}-\frac{1}{2}\right)\erz{j\downarrow}
	\ver{k\downarrow}\left(\bzoMod{k\uparrow}-\frac{1}{2}\right)  
	\\
	\weight{i}{j}{k} &= \left(\!\sqrt2\right)^7 \erz{i\uparrow}\erz{j\downarrow}
	\left(\bzoMod{j\uparrow}-\frac{1}{2}\right)\ver{k\downarrow}
	\left(\bzoMod{k\uparrow}-\frac{1}{2}\right) 
	\\
	\wnine{i}{j}{k} &= \left(\!\sqrt2\right)^9 \erz{i\uparrow}
	\left(\bzoMod{i\downarrow}-\frac{1}{2}\right)\erz{j\downarrow}
	\left(\bzoMod{j\uparrow}-\frac{1}{2}\right)\ver{k\downarrow}
	\left(\bzoMod{k\uparrow}-\frac{1}{2}\right).
\end{alignat}
\end{subequations}
This completes the orthonormal \threePlusBasis{} of operators.

\end{appendix}
\end{widetext}

%\bibliographystyle{aipnum4-1}%was: apsrev
%\bibliography{bibliography_v2}

\begin{thebibliography}{67}%
\makeatletter
\providecommand \@ifxundefined [1]{%
 \@ifx{#1\undefined}
}%
\providecommand \@ifnum [1]{%
 \ifnum #1\expandafter \@firstoftwo
 \else \expandafter \@secondoftwo
 \fi
}%
\providecommand \@ifx [1]{%
 \ifx #1\expandafter \@firstoftwo
 \else \expandafter \@secondoftwo
 \fi
}%
\providecommand \natexlab [1]{#1}%
\providecommand \enquote  [1]{``#1''}%
\providecommand \bibnamefont  [1]{#1}%
\providecommand \bibfnamefont [1]{#1}%
\providecommand \citenamefont [1]{#1}%
\providecommand \href@noop [0]{\@secondoftwo}%
\providecommand \href [0]{\begingroup \@sanitize@url \@href}%
\providecommand \@href[1]{\@@startlink{#1}\@@href}%
\providecommand \@@href[1]{\endgroup#1\@@endlink}%
\providecommand \@sanitize@url [0]{\catcode `\\12\catcode `\$12\catcode
  `\&12\catcode `\#12\catcode `\^12\catcode `\_12\catcode `\%12\relax}%
\providecommand \@@startlink[1]{}%
\providecommand \@@endlink[0]{}%
\providecommand \url  [0]{\begingroup\@sanitize@url \@url }%
\providecommand \@url [1]{\endgroup\@href {#1}{\urlprefix }}%
\providecommand \urlprefix  [0]{URL }%
\providecommand \Eprint [0]{\href }%
\providecommand \doibase [0]{http://dx.doi.org/}%
\providecommand \selectlanguage [0]{\@gobble}%
\providecommand \bibinfo  [0]{\@secondoftwo}%
\providecommand \bibfield  [0]{\@secondoftwo}%
\providecommand \translation [1]{[#1]}%
\providecommand \BibitemOpen [0]{}%
\providecommand \bibitemStop [0]{}%
\providecommand \bibitemNoStop [0]{.\EOS\space}%
\providecommand \EOS [0]{\spacefactor3000\relax}%
\providecommand \BibitemShut  [1]{\csname bibitem#1\endcsname}%
\let\auto@bib@innerbib\@empty
%</preamble>
\bibitem [{\citenamefont {Anderson}\ and\ \citenamefont
  {Kasevich}(1998)}]{Anderson1998}%
  \BibitemOpen
  \bibfield  {author} {\bibinfo {author} {\bibfnamefont {B.~P.}\ \bibnamefont
  {Anderson}}\ and\ \bibinfo {author} {\bibfnamefont {M.~A.}\ \bibnamefont
  {Kasevich}},\ }\href {\doibase 10.1126/science.282.5394.1686} {\bibfield
  {journal} {\bibinfo  {journal} {Science}\ }\textbf {\bibinfo {volume}
  {282}},\ \bibinfo {pages} {1686} (\bibinfo {year} {1998})}\BibitemShut
  {NoStop}%
\bibitem [{\citenamefont {Bloch}(2005)}]{Bloch2005}%
  \BibitemOpen
  \bibfield  {author} {\bibinfo {author} {\bibfnamefont {I.}~\bibnamefont
  {Bloch}},\ }\href {\doibase 10.1038/nphys138} {\bibfield  {journal} {\bibinfo
   {journal} {Nat. Phys.}\ }\textbf {\bibinfo {volume} {1}},\ \bibinfo {pages}
  {23} (\bibinfo {year} {2005})}\BibitemShut {NoStop}%
\bibitem [{\citenamefont {Trotzky}\ \emph {et~al.}(2012)\citenamefont
  {Trotzky}, \citenamefont {Chen}, \citenamefont {Flesch}, \citenamefont
  {McCulloch}, \citenamefont {Schollw{\"{o}}ck}, \citenamefont {Eisert},\ and\
  \citenamefont {Bloch}}]{Trotzky2011}%
  \BibitemOpen
  \bibfield  {author} {\bibinfo {author} {\bibfnamefont {S.}~\bibnamefont
  {Trotzky}}, \bibinfo {author} {\bibfnamefont {Y.~A.}\ \bibnamefont {Chen}},
  \bibinfo {author} {\bibfnamefont {A.}~\bibnamefont {Flesch}}, \bibinfo
  {author} {\bibfnamefont {I.~P.}\ \bibnamefont {McCulloch}}, \bibinfo {author}
  {\bibfnamefont {U.}~\bibnamefont {Schollw{\"{o}}ck}}, \bibinfo {author}
  {\bibfnamefont {J.}~\bibnamefont {Eisert}}, \ and\ \bibinfo {author}
  {\bibfnamefont {I.}~\bibnamefont {Bloch}},\ }\href {\doibase
  10.1038/nphys2232} {\bibfield  {journal} {\bibinfo  {journal} {Nat. Phys.}\
  }\textbf {\bibinfo {volume} {8}},\ \bibinfo {pages} {325} (\bibinfo {year}
  {2012})}\BibitemShut {NoStop}%
\bibitem [{\citenamefont {Greiner}\ \emph {et~al.}(2002)\citenamefont
  {Greiner}, \citenamefont {Mandel}, \citenamefont {Esslinger}, \citenamefont
  {H{\"{a}}nsch},\ and\ \citenamefont {Bloch}}]{Greiner2002}%
  \BibitemOpen
  \bibfield  {author} {\bibinfo {author} {\bibfnamefont {M.}~\bibnamefont
  {Greiner}}, \bibinfo {author} {\bibfnamefont {O.}~\bibnamefont {Mandel}},
  \bibinfo {author} {\bibfnamefont {T.}~\bibnamefont {Esslinger}}, \bibinfo
  {author} {\bibfnamefont {T.~W.}\ \bibnamefont {H{\"{a}}nsch}}, \ and\
  \bibinfo {author} {\bibfnamefont {I.}~\bibnamefont {Bloch}},\ }\href
  {\doibase 10.1038/415039a} {\bibfield  {journal} {\bibinfo  {journal}
  {Nature}\ }\textbf {\bibinfo {volume} {415}},\ \bibinfo {pages} {39}
  (\bibinfo {year} {2002})}\BibitemShut {NoStop}%
\bibitem [{\citenamefont {Goldman}, \citenamefont {Budich},\ and\ \citenamefont
  {Zoller}(2016)}]{Goldman2016}%
  \BibitemOpen
  \bibfield  {author} {\bibinfo {author} {\bibfnamefont {N.}~\bibnamefont
  {Goldman}}, \bibinfo {author} {\bibfnamefont {J.~C.}\ \bibnamefont {Budich}},
  \ and\ \bibinfo {author} {\bibfnamefont {P.}~\bibnamefont {Zoller}},\ }\href
  {\doibase 10.1038/nphys3803} {\bibfield  {journal} {\bibinfo  {journal} {Nat.
  Phys.}\ }\textbf {\bibinfo {volume} {12}},\ \bibinfo {pages} {639} (\bibinfo
  {year} {2016})}\BibitemShut {NoStop}%
\bibitem [{\citenamefont {Axt}\ and\ \citenamefont {Kuhn}(2004)}]{Axt2004}%
  \BibitemOpen
  \bibfield  {author} {\bibinfo {author} {\bibfnamefont {V.~M.}\ \bibnamefont
  {Axt}}\ and\ \bibinfo {author} {\bibfnamefont {T.}~\bibnamefont {Kuhn}},\
  }\href {\doibase 10.1088/0034-4885/67/4/R01} {\bibfield  {journal} {\bibinfo
  {journal} {Reports Prog. Phys.}\ }\textbf {\bibinfo {volume} {67}},\ \bibinfo
  {pages} {433} (\bibinfo {year} {2004})}\BibitemShut {NoStop}%
\bibitem [{\citenamefont {Morawetz}(2004)}]{Morawetz2004}%
  \BibitemOpen
  \bibfield  {author} {\bibinfo {author} {\bibfnamefont {K.}~\bibnamefont
  {Morawetz}},\ }\href {\doibase 10.1007/978-3-662-08990-3} {\emph {\bibinfo
  {title} {{Nonequilibrium Physics at Short Time Scales}}}},\ edited by\
  \bibinfo {editor} {\bibfnamefont {K.}~\bibnamefont {Morawetz}}\ (\bibinfo
  {publisher} {Springer Berlin Heidelberg},\ \bibinfo {address} {Berlin,
  Heidelberg},\ \bibinfo {year} {2004})\BibitemShut {NoStop}%
\bibitem [{\citenamefont {Perfetti}\ \emph {et~al.}(2006)\citenamefont
  {Perfetti}, \citenamefont {Loukakos}, \citenamefont {Lisowski}, \citenamefont
  {Bovensiepen}, \citenamefont {Berger}, \citenamefont {Biermann},
  \citenamefont {Cornaglia}, \citenamefont {Georges},\ and\ \citenamefont
  {Wolf}}]{Perfetti2006}%
  \BibitemOpen
  \bibfield  {author} {\bibinfo {author} {\bibfnamefont {L.}~\bibnamefont
  {Perfetti}}, \bibinfo {author} {\bibfnamefont {P.~A.}\ \bibnamefont
  {Loukakos}}, \bibinfo {author} {\bibfnamefont {M.}~\bibnamefont {Lisowski}},
  \bibinfo {author} {\bibfnamefont {U.}~\bibnamefont {Bovensiepen}}, \bibinfo
  {author} {\bibfnamefont {H.}~\bibnamefont {Berger}}, \bibinfo {author}
  {\bibfnamefont {S.}~\bibnamefont {Biermann}}, \bibinfo {author}
  {\bibfnamefont {P.~S.}\ \bibnamefont {Cornaglia}}, \bibinfo {author}
  {\bibfnamefont {A.}~\bibnamefont {Georges}}, \ and\ \bibinfo {author}
  {\bibfnamefont {M.}~\bibnamefont {Wolf}},\ }\href {\doibase
  10.1103/PhysRevLett.97.067402} {\bibfield  {journal} {\bibinfo  {journal}
  {Phys. Rev. Lett.}\ }\textbf {\bibinfo {volume} {97}},\ \bibinfo {pages}
  {067402} (\bibinfo {year} {2006})}\BibitemShut {NoStop}%
\bibitem [{\citenamefont {St{\"{o}}ferle}\ \emph {et~al.}(2004)\citenamefont
  {St{\"{o}}ferle}, \citenamefont {Moritz}, \citenamefont {Schori},
  \citenamefont {K{\"{o}}hl},\ and\ \citenamefont {Esslinger}}]{Stoferle2004}%
  \BibitemOpen
  \bibfield  {author} {\bibinfo {author} {\bibfnamefont {T.}~\bibnamefont
  {St{\"{o}}ferle}}, \bibinfo {author} {\bibfnamefont {H.}~\bibnamefont
  {Moritz}}, \bibinfo {author} {\bibfnamefont {C.}~\bibnamefont {Schori}},
  \bibinfo {author} {\bibfnamefont {M.}~\bibnamefont {K{\"{o}}hl}}, \ and\
  \bibinfo {author} {\bibfnamefont {T.}~\bibnamefont {Esslinger}},\ }\href
  {\doibase 10.1103/PhysRevLett.92.130403} {\bibfield  {journal} {\bibinfo
  {journal} {Phys. Rev. Lett.}\ }\textbf {\bibinfo {volume} {92}},\ \bibinfo
  {pages} {130403} (\bibinfo {year} {2004})}\BibitemShut {NoStop}%
\bibitem [{\citenamefont {Sherson}\ \emph {et~al.}(2010)\citenamefont
  {Sherson}, \citenamefont {Weitenberg}, \citenamefont {Endres}, \citenamefont
  {Cheneau}, \citenamefont {Bloch},\ and\ \citenamefont {Kuhr}}]{Sherson2010}%
  \BibitemOpen
  \bibfield  {author} {\bibinfo {author} {\bibfnamefont {J.~F.}\ \bibnamefont
  {Sherson}}, \bibinfo {author} {\bibfnamefont {C.}~\bibnamefont {Weitenberg}},
  \bibinfo {author} {\bibfnamefont {M.}~\bibnamefont {Endres}}, \bibinfo
  {author} {\bibfnamefont {M.}~\bibnamefont {Cheneau}}, \bibinfo {author}
  {\bibfnamefont {I.}~\bibnamefont {Bloch}}, \ and\ \bibinfo {author}
  {\bibfnamefont {S.}~\bibnamefont {Kuhr}},\ }\href {\doibase
  10.1038/nature09378} {\bibfield  {journal} {\bibinfo  {journal} {Nature}\
  }\textbf {\bibinfo {volume} {467}},\ \bibinfo {pages} {68} (\bibinfo {year}
  {2010})}\BibitemShut {NoStop}%
\bibitem [{\citenamefont {Sanders}, \citenamefont {Mintert},\ and\
  \citenamefont {Heller}(2010)}]{Sanders2010}%
  \BibitemOpen
  \bibfield  {author} {\bibinfo {author} {\bibfnamefont {S.~N.}\ \bibnamefont
  {Sanders}}, \bibinfo {author} {\bibfnamefont {F.}~\bibnamefont {Mintert}}, \
  and\ \bibinfo {author} {\bibfnamefont {E.~J.}\ \bibnamefont {Heller}},\
  }\href {\doibase 10.1103/PhysRevLett.105.035301} {\bibfield  {journal}
  {\bibinfo  {journal} {Phys. Rev. Lett.}\ }\textbf {\bibinfo {volume} {105}},\
  \bibinfo {pages} {035301} (\bibinfo {year} {2010})}\BibitemShut {NoStop}%
\bibitem [{\citenamefont {Mayer}, \citenamefont {Rodriguez},\ and\
  \citenamefont {Buchleitner}(2014)}]{Mayer2014}%
  \BibitemOpen
  \bibfield  {author} {\bibinfo {author} {\bibfnamefont {K.}~\bibnamefont
  {Mayer}}, \bibinfo {author} {\bibfnamefont {A.}~\bibnamefont {Rodriguez}}, \
  and\ \bibinfo {author} {\bibfnamefont {A.}~\bibnamefont {Buchleitner}},\
  }\href {\doibase 10.1103/PhysRevA.90.023629} {\bibfield  {journal} {\bibinfo
  {journal} {Phys. Rev. A}\ }\textbf {\bibinfo {volume} {90}},\ \bibinfo
  {pages} {023629} (\bibinfo {year} {2014})}\BibitemShut {NoStop}%
\bibitem [{\citenamefont {Mekhov}, \citenamefont {Maschler},\ and\
  \citenamefont {Ritsch}(2007)}]{Mekhov2007}%
  \BibitemOpen
  \bibfield  {author} {\bibinfo {author} {\bibfnamefont {I.~B.}\ \bibnamefont
  {Mekhov}}, \bibinfo {author} {\bibfnamefont {C.}~\bibnamefont {Maschler}}, \
  and\ \bibinfo {author} {\bibfnamefont {H.}~\bibnamefont {Ritsch}},\ }\href
  {\doibase 10.1038/nphys571} {\bibfield  {journal} {\bibinfo  {journal} {Nat.
  Phys.}\ }\textbf {\bibinfo {volume} {3}},\ \bibinfo {pages} {319} (\bibinfo
  {year} {2007})}\BibitemShut {NoStop}%
\bibitem [{\citenamefont {ten Brinke}\ and\ \citenamefont
  {Sch{\"{u}}tzhold}(2015)}]{TenBrinke2015}%
  \BibitemOpen
  \bibfield  {author} {\bibinfo {author} {\bibfnamefont {N.}~\bibnamefont {ten
  Brinke}}\ and\ \bibinfo {author} {\bibfnamefont {R.}~\bibnamefont
  {Sch{\"{u}}tzhold}},\ }\href {\doibase 10.1103/PhysRevA.92.013617} {\bibfield
   {journal} {\bibinfo  {journal} {Phys. Rev. A}\ }\textbf {\bibinfo {volume}
  {92}},\ \bibinfo {pages} {013617} (\bibinfo {year} {2015})}\BibitemShut
  {NoStop}%
\bibitem [{\citenamefont {Strohmaier}\ \emph {et~al.}(2010)\citenamefont
  {Strohmaier}, \citenamefont {Greif}, \citenamefont {J{\"{o}}rdens},
  \citenamefont {Tarruell}, \citenamefont {Moritz}, \citenamefont {Esslinger},
  \citenamefont {Sensarma}, \citenamefont {Pekker}, \citenamefont {Altman},\
  and\ \citenamefont {Demler}}]{Strohmaier2010}%
  \BibitemOpen
  \bibfield  {author} {\bibinfo {author} {\bibfnamefont {N.}~\bibnamefont
  {Strohmaier}}, \bibinfo {author} {\bibfnamefont {D.}~\bibnamefont {Greif}},
  \bibinfo {author} {\bibfnamefont {R.}~\bibnamefont {J{\"{o}}rdens}}, \bibinfo
  {author} {\bibfnamefont {L.}~\bibnamefont {Tarruell}}, \bibinfo {author}
  {\bibfnamefont {H.}~\bibnamefont {Moritz}}, \bibinfo {author} {\bibfnamefont
  {T.}~\bibnamefont {Esslinger}}, \bibinfo {author} {\bibfnamefont
  {R.}~\bibnamefont {Sensarma}}, \bibinfo {author} {\bibfnamefont
  {D.}~\bibnamefont {Pekker}}, \bibinfo {author} {\bibfnamefont
  {E.}~\bibnamefont {Altman}}, \ and\ \bibinfo {author} {\bibfnamefont
  {E.}~\bibnamefont {Demler}},\ }\href {\doibase
  10.1103/PhysRevLett.104.080401} {\bibfield  {journal} {\bibinfo  {journal}
  {Phys. Rev. Lett.}\ }\textbf {\bibinfo {volume} {104}},\ \bibinfo {pages}
  {080401} (\bibinfo {year} {2010})}\BibitemShut {NoStop}%
\bibitem [{\citenamefont {Lenar\v{c}i\v{c}}\ and\ \citenamefont
  {Prelov\v{s}ek}(2013)}]{lenar13}%
  \BibitemOpen
  \bibfield  {author} {\bibinfo {author} {\bibfnamefont {Z.}~\bibnamefont
  {Lenar\v{c}i\v{c}}}\ and\ \bibinfo {author} {\bibfnamefont {P.}~\bibnamefont
  {Prelov\v{s}ek}},\ }\href@noop {} {\bibfield  {journal} {\bibinfo  {journal}
  {Phys. Rev. Lett.}\ }\textbf {\bibinfo {volume} {111}},\ \bibinfo {pages}
  {016401} (\bibinfo {year} {2013})}\BibitemShut {NoStop}%
\bibitem [{\citenamefont {Mattuck}(1976)}]{Mattuck1976}%
  \BibitemOpen
  \bibfield  {author} {\bibinfo {author} {\bibfnamefont {R.~D.}\ \bibnamefont
  {Mattuck}},\ }\href@noop {} {\emph {\bibinfo {title} {{A guide to Feynman
  diagrams in the many-body problem}}}}\ (\bibinfo  {publisher} {Dover
  Publications},\ \bibinfo {address} {New York},\ \bibinfo {year}
  {1976})\BibitemShut {NoStop}%
\bibitem [{\citenamefont {L{\"{a}}uchli}\ and\ \citenamefont
  {Kollath}(2008)}]{Lauchli2008}%
  \BibitemOpen
  \bibfield  {author} {\bibinfo {author} {\bibfnamefont {A.~M.}\ \bibnamefont
  {L{\"{a}}uchli}}\ and\ \bibinfo {author} {\bibfnamefont {C.}~\bibnamefont
  {Kollath}},\ }\href {\doibase 10.1088/1742-5468/2008/05/P05018} {\bibfield
  {journal} {\bibinfo  {journal} {J. Stat. Mech. Theory Exp.}\ }\textbf
  {\bibinfo {volume} {2008}},\ \bibinfo {pages} {P05018} (\bibinfo {year}
  {2008})}\BibitemShut {NoStop}%
\bibitem [{\citenamefont {Chen}\ \emph {et~al.}(2011)\citenamefont {Chen},
  \citenamefont {White}, \citenamefont {Borries},\ and\ \citenamefont
  {DeMarco}}]{Chen2011}%
  \BibitemOpen
  \bibfield  {author} {\bibinfo {author} {\bibfnamefont {D.}~\bibnamefont
  {Chen}}, \bibinfo {author} {\bibfnamefont {M.}~\bibnamefont {White}},
  \bibinfo {author} {\bibfnamefont {C.}~\bibnamefont {Borries}}, \ and\
  \bibinfo {author} {\bibfnamefont {B.}~\bibnamefont {DeMarco}},\ }\href
  {\doibase 10.1103/PhysRevLett.106.235304} {\bibfield  {journal} {\bibinfo
  {journal} {Phys. Rev. Lett.}\ }\textbf {\bibinfo {volume} {106}},\ \bibinfo
  {pages} {235304} (\bibinfo {year} {2011})}\BibitemShut {NoStop}%
\bibitem [{\citenamefont {Langen}\ \emph {et~al.}(2013)\citenamefont {Langen},
  \citenamefont {Geiger}, \citenamefont {Kuhnert}, \citenamefont {Rauer},\ and\
  \citenamefont {Schmiedmayer}}]{Langen2013}%
  \BibitemOpen
  \bibfield  {author} {\bibinfo {author} {\bibfnamefont {T.}~\bibnamefont
  {Langen}}, \bibinfo {author} {\bibfnamefont {R.}~\bibnamefont {Geiger}},
  \bibinfo {author} {\bibfnamefont {M.}~\bibnamefont {Kuhnert}}, \bibinfo
  {author} {\bibfnamefont {B.}~\bibnamefont {Rauer}}, \ and\ \bibinfo {author}
  {\bibfnamefont {J.}~\bibnamefont {Schmiedmayer}},\ }\href {\doibase
  10.1038/nphys2739} {\bibfield  {journal} {\bibinfo  {journal} {Nat. Phys.}\
  }\textbf {\bibinfo {volume} {9}},\ \bibinfo {pages} {640} (\bibinfo {year}
  {2013})}\BibitemShut {NoStop}%
\bibitem [{\citenamefont {Cazalilla}(2006)}]{cazal06}%
  \BibitemOpen
  \bibfield  {author} {\bibinfo {author} {\bibfnamefont {M.~A.}\ \bibnamefont
  {Cazalilla}},\ }\href@noop {} {\bibfield  {journal} {\bibinfo  {journal}
  {Phys. Rev. Lett.}\ }\textbf {\bibinfo {volume} {97}},\ \bibinfo {pages}
  {156403} (\bibinfo {year} {2006})}\BibitemShut {NoStop}%
\bibitem [{\citenamefont {Barthel}\ and\ \citenamefont
  {Schollw\"ock}(2008)}]{barth08}%
  \BibitemOpen
  \bibfield  {author} {\bibinfo {author} {\bibfnamefont {T.}~\bibnamefont
  {Barthel}}\ and\ \bibinfo {author} {\bibfnamefont {U.}~\bibnamefont
  {Schollw\"ock}},\ }\href@noop {} {\bibfield  {journal} {\bibinfo  {journal}
  {Phys. Rev. Lett.}\ }\textbf {\bibinfo {volume} {100}},\ \bibinfo {pages}
  {100601} (\bibinfo {year} {2008})}\BibitemShut {NoStop}%
\bibitem [{\citenamefont {Calabrese}, \citenamefont {Essler},\ and\
  \citenamefont {Fagotti}(2011)}]{Calabrese2011}%
  \BibitemOpen
  \bibfield  {author} {\bibinfo {author} {\bibfnamefont {P.}~\bibnamefont
  {Calabrese}}, \bibinfo {author} {\bibfnamefont {F.~H.~L.}\ \bibnamefont
  {Essler}}, \ and\ \bibinfo {author} {\bibfnamefont {M.}~\bibnamefont
  {Fagotti}},\ }\href {\doibase 10.1103/PhysRevLett.106.227203} {\bibfield
  {journal} {\bibinfo  {journal} {Phys. Rev. Lett.}\ }\textbf {\bibinfo
  {volume} {106}},\ \bibinfo {pages} {227203} (\bibinfo {year}
  {2011})}\BibitemShut {NoStop}%
\bibitem [{\citenamefont {Calabrese}, \citenamefont {Essler},\ and\
  \citenamefont {Fagotti}(2012{\natexlab{a}})}]{Calabrese2012a}%
  \BibitemOpen
  \bibfield  {author} {\bibinfo {author} {\bibfnamefont {P.}~\bibnamefont
  {Calabrese}}, \bibinfo {author} {\bibfnamefont {F.~H.~L.}\ \bibnamefont
  {Essler}}, \ and\ \bibinfo {author} {\bibfnamefont {M.}~\bibnamefont
  {Fagotti}},\ }\href {\doibase 10.1088/1742-5468/2012/07/P07022} {\bibfield
  {journal} {\bibinfo  {journal} {J. Stat. Mech. Theory Exp.}\ }\textbf
  {\bibinfo {volume} {2012}},\ \bibinfo {pages} {P07022} (\bibinfo {year}
  {2012}{\natexlab{a}})}\BibitemShut {NoStop}%
\bibitem [{\citenamefont {Calabrese}, \citenamefont {Essler},\ and\
  \citenamefont {Fagotti}(2012{\natexlab{b}})}]{calab12b}%
  \BibitemOpen
  \bibfield  {author} {\bibinfo {author} {\bibfnamefont {P.}~\bibnamefont
  {Calabrese}}, \bibinfo {author} {\bibfnamefont {F.~H.~L.}\ \bibnamefont
  {Essler}}, \ and\ \bibinfo {author} {\bibfnamefont {M.}~\bibnamefont
  {Fagotti}},\ }\href@noop {} {\bibfield  {journal} {\bibinfo  {journal} {J.
  Stat. Mech. Theory Exp.}\ ,\ \bibinfo {pages} {P07022}} (\bibinfo {year}
  {2012}{\natexlab{b}})}\BibitemShut {NoStop}%
\bibitem [{\citenamefont {Caux}\ and\ \citenamefont {Essler}(2013)}]{Caux2013}%
  \BibitemOpen
  \bibfield  {author} {\bibinfo {author} {\bibfnamefont {J.-S.}\ \bibnamefont
  {Caux}}\ and\ \bibinfo {author} {\bibfnamefont {F.~H.~L.}\ \bibnamefont
  {Essler}},\ }\href {\doibase 10.1103/PhysRevLett.110.257203} {\bibfield
  {journal} {\bibinfo  {journal} {Phys. Rev. Lett.}\ }\textbf {\bibinfo
  {volume} {110}},\ \bibinfo {pages} {257203} (\bibinfo {year}
  {2013})}\BibitemShut {NoStop}%
\bibitem [{\citenamefont {Rigol}(2009)}]{Rigol2009}%
  \BibitemOpen
  \bibfield  {author} {\bibinfo {author} {\bibfnamefont {M.}~\bibnamefont
  {Rigol}},\ }\href {\doibase 10.1103/PhysRevA.80.053607} {\bibfield  {journal}
  {\bibinfo  {journal} {Phys. Rev. A}\ }\textbf {\bibinfo {volume} {80}},\
  \bibinfo {pages} {053607} (\bibinfo {year} {2009})}\BibitemShut {NoStop}%
\bibitem [{\citenamefont {Eckstein}, \citenamefont {Kollar},\ and\
  \citenamefont {Werner}(2009)}]{Eckstein2009}%
  \BibitemOpen
  \bibfield  {author} {\bibinfo {author} {\bibfnamefont {M.}~\bibnamefont
  {Eckstein}}, \bibinfo {author} {\bibfnamefont {M.}~\bibnamefont {Kollar}}, \
  and\ \bibinfo {author} {\bibfnamefont {P.}~\bibnamefont {Werner}},\ }\href
  {\doibase 10.1103/PhysRevLett.103.056403} {\bibfield  {journal} {\bibinfo
  {journal} {Phys. Rev. Lett.}\ }\textbf {\bibinfo {volume} {103}},\ \bibinfo
  {pages} {056403} (\bibinfo {year} {2009})}\BibitemShut {NoStop}%
\bibitem [{\citenamefont {Aoki}\ \emph {et~al.}(2014)\citenamefont {Aoki},
  \citenamefont {Tsuji}, \citenamefont {Eckstein}, \citenamefont {Kollar},
  \citenamefont {Oka},\ and\ \citenamefont {Werner}}]{Aoki2014}%
  \BibitemOpen
  \bibfield  {author} {\bibinfo {author} {\bibfnamefont {H.}~\bibnamefont
  {Aoki}}, \bibinfo {author} {\bibfnamefont {N.}~\bibnamefont {Tsuji}},
  \bibinfo {author} {\bibfnamefont {M.}~\bibnamefont {Eckstein}}, \bibinfo
  {author} {\bibfnamefont {M.}~\bibnamefont {Kollar}}, \bibinfo {author}
  {\bibfnamefont {T.}~\bibnamefont {Oka}}, \ and\ \bibinfo {author}
  {\bibfnamefont {P.}~\bibnamefont {Werner}},\ }\href {\doibase
  10.1103/RevModPhys.86.779} {\bibfield  {journal} {\bibinfo  {journal} {Rev.
  Mod. Phys.}\ }\textbf {\bibinfo {volume} {86}},\ \bibinfo {pages} {779}
  (\bibinfo {year} {2014})}\BibitemShut {NoStop}%
\bibitem [{\citenamefont {Navez}\ and\ \citenamefont
  {Sch{\"{u}}tzhold}(2010)}]{Navez2010}%
  \BibitemOpen
  \bibfield  {author} {\bibinfo {author} {\bibfnamefont {P.}~\bibnamefont
  {Navez}}\ and\ \bibinfo {author} {\bibfnamefont {R.}~\bibnamefont
  {Sch{\"{u}}tzhold}},\ }\href {\doibase 10.1103/PhysRevA.82.063603} {\bibfield
   {journal} {\bibinfo  {journal} {Phys. Rev. A}\ }\textbf {\bibinfo {volume}
  {82}},\ \bibinfo {pages} {063603} (\bibinfo {year} {2010})}\BibitemShut
  {NoStop}%
\bibitem [{\citenamefont {Krutitsky}\ \emph {et~al.}(2014)\citenamefont
  {Krutitsky}, \citenamefont {Navez}, \citenamefont {Queisser},\ and\
  \citenamefont {Sch{\"{u}}tzhold}}]{Krutitsky2014}%
  \BibitemOpen
  \bibfield  {author} {\bibinfo {author} {\bibfnamefont {K.~V.}\ \bibnamefont
  {Krutitsky}}, \bibinfo {author} {\bibfnamefont {P.}~\bibnamefont {Navez}},
  \bibinfo {author} {\bibfnamefont {F.}~\bibnamefont {Queisser}}, \ and\
  \bibinfo {author} {\bibfnamefont {R.}~\bibnamefont {Sch{\"{u}}tzhold}},\
  }\href {\doibase 10.1140/epjqt12} {\bibfield  {journal} {\bibinfo  {journal}
  {EPJ Quantum Technol.}\ }\textbf {\bibinfo {volume} {1}},\ \bibinfo {pages}
  {12} (\bibinfo {year} {2014})}\BibitemShut {NoStop}%
\bibitem [{\citenamefont {White}\ and\ \citenamefont
  {Feiguin}(2004)}]{white04a}%
  \BibitemOpen
  \bibfield  {author} {\bibinfo {author} {\bibfnamefont {S.~R.}\ \bibnamefont
  {White}}\ and\ \bibinfo {author} {\bibfnamefont {A.~E.}\ \bibnamefont
  {Feiguin}},\ }\href@noop {} {\bibfield  {journal} {\bibinfo  {journal} {Phys.
  Rev. Lett.}\ }\textbf {\bibinfo {volume} {93}},\ \bibinfo {pages} {076401}
  (\bibinfo {year} {2004})}\BibitemShut {NoStop}%
\bibitem [{\citenamefont {Daley}\ \emph {et~al.}(2004)\citenamefont {Daley},
  \citenamefont {Kollath}, \citenamefont {Schollw{\"{o}}ck},\ and\
  \citenamefont {Vidal}}]{Daley2004}%
  \BibitemOpen
  \bibfield  {author} {\bibinfo {author} {\bibfnamefont {A.~J.}\ \bibnamefont
  {Daley}}, \bibinfo {author} {\bibfnamefont {C.}~\bibnamefont {Kollath}},
  \bibinfo {author} {\bibfnamefont {U.}~\bibnamefont {Schollw{\"{o}}ck}}, \
  and\ \bibinfo {author} {\bibfnamefont {G.}~\bibnamefont {Vidal}},\ }\href
  {\doibase 10.1088/1742-5468/2004/04/P04005} {\bibfield  {journal} {\bibinfo
  {journal} {J. Stat. Mech. Theory Exp.}\ }\textbf {\bibinfo {volume} {2004}},\
  \bibinfo {pages} {P04005} (\bibinfo {year} {2004})}\BibitemShut {NoStop}%
\bibitem [{\citenamefont {Batrouni}\ \emph {et~al.}(2005)\citenamefont
  {Batrouni}, \citenamefont {Assaad}, \citenamefont {Scalettar},\ and\
  \citenamefont {Denteneer}}]{Batrouni2005}%
  \BibitemOpen
  \bibfield  {author} {\bibinfo {author} {\bibfnamefont {G.~G.}\ \bibnamefont
  {Batrouni}}, \bibinfo {author} {\bibfnamefont {F.~F.}\ \bibnamefont
  {Assaad}}, \bibinfo {author} {\bibfnamefont {R.~T.}\ \bibnamefont
  {Scalettar}}, \ and\ \bibinfo {author} {\bibfnamefont {P.~J.~H.}\
  \bibnamefont {Denteneer}},\ }\href {\doibase 10.1103/PhysRevA.72.031601}
  {\bibfield  {journal} {\bibinfo  {journal} {Phys. Rev. A}\ }\textbf {\bibinfo
  {volume} {72}},\ \bibinfo {pages} {031601(R)} (\bibinfo {year}
  {2005})}\BibitemShut {NoStop}%
\bibitem [{\citenamefont {Goth}\ and\ \citenamefont {Assaad}(2012)}]{Goth2012}%
  \BibitemOpen
  \bibfield  {author} {\bibinfo {author} {\bibfnamefont {F.}~\bibnamefont
  {Goth}}\ and\ \bibinfo {author} {\bibfnamefont {F.~F.}\ \bibnamefont
  {Assaad}},\ }\href {\doibase 10.1103/PhysRevB.85.085129} {\bibfield
  {journal} {\bibinfo  {journal} {Phys. Rev. B}\ }\textbf {\bibinfo {volume}
  {85}},\ \bibinfo {pages} {085129} (\bibinfo {year} {2012})}\BibitemShut
  {NoStop}%
\bibitem [{\citenamefont {Schir{\'{o}}}\ and\ \citenamefont
  {Fabrizio}(2010)}]{Schiro2010}%
  \BibitemOpen
  \bibfield  {author} {\bibinfo {author} {\bibfnamefont {M.}~\bibnamefont
  {Schir{\'{o}}}}\ and\ \bibinfo {author} {\bibfnamefont {M.}~\bibnamefont
  {Fabrizio}},\ }\href {\doibase 10.1103/PhysRevLett.105.076401} {\bibfield
  {journal} {\bibinfo  {journal} {Phys. Rev. Lett.}\ }\textbf {\bibinfo
  {volume} {105}},\ \bibinfo {pages} {076401} (\bibinfo {year}
  {2010})}\BibitemShut {NoStop}%
\bibitem [{\citenamefont {Schir{\'{o}}}\ and\ \citenamefont
  {Fabrizio}(2011)}]{Schiro2011}%
  \BibitemOpen
  \bibfield  {author} {\bibinfo {author} {\bibfnamefont {M.}~\bibnamefont
  {Schir{\'{o}}}}\ and\ \bibinfo {author} {\bibfnamefont {M.}~\bibnamefont
  {Fabrizio}},\ }\href {\doibase 10.1103/PhysRevB.83.165105} {\bibfield
  {journal} {\bibinfo  {journal} {Phys. Rev. B}\ }\textbf {\bibinfo {volume}
  {83}},\ \bibinfo {pages} {165105} (\bibinfo {year} {2011})}\BibitemShut
  {NoStop}%
\bibitem [{\citenamefont {Ido}, \citenamefont {Ohgoe},\ and\ \citenamefont
  {Imada}(2015)}]{ido15}%
  \BibitemOpen
  \bibfield  {author} {\bibinfo {author} {\bibfnamefont {K.}~\bibnamefont
  {Ido}}, \bibinfo {author} {\bibfnamefont {T.}~\bibnamefont {Ohgoe}}, \ and\
  \bibinfo {author} {\bibfnamefont {M.}~\bibnamefont {Imada}},\ }\href@noop {}
  {\bibfield  {journal} {\bibinfo  {journal} {Phys. Rev. B}\ }\textbf {\bibinfo
  {volume} {92}},\ \bibinfo {pages} {245106} (\bibinfo {year}
  {2015})}\BibitemShut {NoStop}%
\bibitem [{\citenamefont {Moeckel}\ and\ \citenamefont
  {Kehrein}(2008)}]{Moeckel2008}%
  \BibitemOpen
  \bibfield  {author} {\bibinfo {author} {\bibfnamefont {M.}~\bibnamefont
  {Moeckel}}\ and\ \bibinfo {author} {\bibfnamefont {S.}~\bibnamefont
  {Kehrein}},\ }\href {\doibase 10.1103/PhysRevLett.100.175702} {\bibfield
  {journal} {\bibinfo  {journal} {Phys. Rev. Lett.}\ }\textbf {\bibinfo
  {volume} {100}},\ \bibinfo {pages} {175702} (\bibinfo {year}
  {2008})}\BibitemShut {NoStop}%
\bibitem [{\citenamefont {Moeckel}\ and\ \citenamefont
  {Kehrein}(2009)}]{Moeckel2009}%
  \BibitemOpen
  \bibfield  {author} {\bibinfo {author} {\bibfnamefont {M.}~\bibnamefont
  {Moeckel}}\ and\ \bibinfo {author} {\bibfnamefont {S.}~\bibnamefont
  {Kehrein}},\ }\href {\doibase 10.1016/j.aop.2009.03.009} {\bibfield
  {journal} {\bibinfo  {journal} {Ann. Phys. (N. Y).}\ }\textbf {\bibinfo
  {volume} {324}},\ \bibinfo {pages} {2146} (\bibinfo {year}
  {2009})}\BibitemShut {NoStop}%
\bibitem [{\citenamefont {Sabio}\ and\ \citenamefont
  {Kehrein}(2010)}]{Sabio2010}%
  \BibitemOpen
  \bibfield  {author} {\bibinfo {author} {\bibfnamefont {J.}~\bibnamefont
  {Sabio}}\ and\ \bibinfo {author} {\bibfnamefont {S.}~\bibnamefont
  {Kehrein}},\ }\href {\doibase 10.1088/1367-2630/12/5/055008} {\bibfield
  {journal} {\bibinfo  {journal} {New J. Phys.}\ }\textbf {\bibinfo {volume}
  {12}},\ \bibinfo {pages} {055008} (\bibinfo {year} {2010})}\BibitemShut
  {NoStop}%
\bibitem [{\citenamefont {Uhrig}(2009)}]{Uhrig2009}%
  \BibitemOpen
  \bibfield  {author} {\bibinfo {author} {\bibfnamefont {G.~S.}\ \bibnamefont
  {Uhrig}},\ }\href {\doibase 10.1103/PhysRevA.80.061602} {\bibfield  {journal}
  {\bibinfo  {journal} {Phys. Rev. A}\ }\textbf {\bibinfo {volume} {80}},\
  \bibinfo {pages} {061602(R)} (\bibinfo {year} {2009})}\BibitemShut {NoStop}%
\bibitem [{\citenamefont {Hamerla}\ and\ \citenamefont
  {Uhrig}(2013{\natexlab{a}})}]{Hamerla2013}%
  \BibitemOpen
  \bibfield  {author} {\bibinfo {author} {\bibfnamefont {S.~A.}\ \bibnamefont
  {Hamerla}}\ and\ \bibinfo {author} {\bibfnamefont {G.~S.}\ \bibnamefont
  {Uhrig}},\ }\href {\doibase 10.1103/PhysRevB.87.064304} {\bibfield  {journal}
  {\bibinfo  {journal} {Phys. Rev. B}\ }\textbf {\bibinfo {volume} {87}},\
  \bibinfo {pages} {064304} (\bibinfo {year} {2013}{\natexlab{a}})}\BibitemShut
  {NoStop}%
\bibitem [{\citenamefont {Hamerla}\ and\ \citenamefont
  {Uhrig}(2013{\natexlab{b}})}]{Hamerla2013a}%
  \BibitemOpen
  \bibfield  {author} {\bibinfo {author} {\bibfnamefont {S.~A.}\ \bibnamefont
  {Hamerla}}\ and\ \bibinfo {author} {\bibfnamefont {G.~S.}\ \bibnamefont
  {Uhrig}},\ }\href {\doibase 10.1088/1367-2630/15/7/073012} {\bibfield
  {journal} {\bibinfo  {journal} {New J. Phys.}\ }\textbf {\bibinfo {volume}
  {15}},\ \bibinfo {pages} {073012} (\bibinfo {year}
  {2013}{\natexlab{b}})}\BibitemShut {NoStop}%
\bibitem [{\citenamefont {Hamerla}\ and\ \citenamefont
  {Uhrig}(2014)}]{Hamerla2014}%
  \BibitemOpen
  \bibfield  {author} {\bibinfo {author} {\bibfnamefont {S.~A.}\ \bibnamefont
  {Hamerla}}\ and\ \bibinfo {author} {\bibfnamefont {G.~S.}\ \bibnamefont
  {Uhrig}},\ }\href {\doibase 10.1103/PhysRevB.89.104301} {\bibfield  {journal}
  {\bibinfo  {journal} {Phys. Rev. B}\ }\textbf {\bibinfo {volume} {89}},\
  \bibinfo {pages} {104301} (\bibinfo {year} {2014})}\BibitemShut {NoStop}%
\bibitem [{\citenamefont {Kalthoff}\ \emph {et~al.}(2017)\citenamefont
  {Kalthoff}, \citenamefont {Keim}, \citenamefont {Krull},\ and\ \citenamefont
  {Uhrig}}]{Kalthoff2017}%
  \BibitemOpen
  \bibfield  {author} {\bibinfo {author} {\bibfnamefont {M.}~\bibnamefont
  {Kalthoff}}, \bibinfo {author} {\bibfnamefont {F.}~\bibnamefont {Keim}},
  \bibinfo {author} {\bibfnamefont {H.}~\bibnamefont {Krull}}, \ and\ \bibinfo
  {author} {\bibfnamefont {G.~S.}\ \bibnamefont {Uhrig}},\ }\href {\doibase
  10.1140/epjb/e2017-80063-2} {\bibfield  {journal} {\bibinfo  {journal} {Eur.
  Phys. J. B}\ }\textbf {\bibinfo {volume} {90}},\ \bibinfo {pages} {97}
  (\bibinfo {year} {2017})}\BibitemShut {NoStop}%
\bibitem [{\citenamefont {R{\"{o}}hrig}\ \emph {et~al.}(2018)\citenamefont
  {R{\"{o}}hrig}, \citenamefont {Schering}, \citenamefont {Gravert},\ and\
  \citenamefont {Uhrig}}]{Rohrig2017}%
  \BibitemOpen
  \bibfield  {author} {\bibinfo {author} {\bibfnamefont {R.}~\bibnamefont
  {R{\"{o}}hrig}}, \bibinfo {author} {\bibfnamefont {P.}~\bibnamefont
  {Schering}}, \bibinfo {author} {\bibfnamefont {L.~B.}\ \bibnamefont
  {Gravert}}, \ and\ \bibinfo {author} {\bibfnamefont {G.~S.}\ \bibnamefont
  {Uhrig}},\ }\href {http://arxiv.org/abs/1711.08919} {\bibfield  {journal}
  {\bibinfo  {journal} {Phys. Rev. B (in press)}\ } (\bibinfo {year} {2018})},\
  \Eprint {http://arxiv.org/abs/1711.08919} {arXiv:1711.08919} \BibitemShut
  {NoStop}%
\bibitem [{\citenamefont {Lieb}\ and\ \citenamefont {Wu}(1968)}]{Lieb1968}%
  \BibitemOpen
  \bibfield  {author} {\bibinfo {author} {\bibfnamefont {E.~H.}\ \bibnamefont
  {Lieb}}\ and\ \bibinfo {author} {\bibfnamefont {F.~Y.}\ \bibnamefont {Wu}},\
  }\href {\doibase 10.1103/PhysRevLett.20.1445} {\bibfield  {journal} {\bibinfo
   {journal} {Phys. Rev. Lett.}\ }\textbf {\bibinfo {volume} {20}},\ \bibinfo
  {pages} {1445} (\bibinfo {year} {1968})}\BibitemShut {NoStop}%
\bibitem [{\citenamefont {Faddeev}(2005)}]{Essler2005}%
  \BibitemOpen
  \bibfield  {author} {\bibinfo {author} {\bibfnamefont {L.~D.}\ \bibnamefont
  {Faddeev}},\ }\href {\doibase 10.1017/CBO9780511534843.004} {\emph {\bibinfo
  {title} {{The Bethe ansatz}}}}\ (\bibinfo  {publisher} {Cambridge University
  Press},\ \bibinfo {address} {Cambridge},\ \bibinfo {year} {2005})\BibitemShut
  {NoStop}%
\bibitem [{\citenamefont {Hubbard}(1963)}]{Hubbard1963}%
  \BibitemOpen
  \bibfield  {author} {\bibinfo {author} {\bibfnamefont {J.}~\bibnamefont
  {Hubbard}},\ }\href {\doibase 10.1098/rspa.1963.0204} {\bibfield  {journal}
  {\bibinfo  {journal} {Proc. R. Soc. A Math. Phys. Eng. Sci.}\ }\textbf
  {\bibinfo {volume} {276}},\ \bibinfo {pages} {238} (\bibinfo {year}
  {1963})}\BibitemShut {NoStop}%
\bibitem [{\citenamefont {Kanamori}(1963)}]{Kanamori1963}%
  \BibitemOpen
  \bibfield  {author} {\bibinfo {author} {\bibfnamefont {J.}~\bibnamefont
  {Kanamori}},\ }\href {\doibase 10.1143/PTP.30.275} {\bibfield  {journal}
  {\bibinfo  {journal} {Prog. Theor. Phys.}\ }\textbf {\bibinfo {volume}
  {30}},\ \bibinfo {pages} {275} (\bibinfo {year} {1963})}\BibitemShut
  {NoStop}%
\bibitem [{\citenamefont {Gutzwiller}(1963)}]{gutzw63}%
  \BibitemOpen
  \bibfield  {author} {\bibinfo {author} {\bibfnamefont {M.~C.}\ \bibnamefont
  {Gutzwiller}},\ }\href@noop {} {\bibfield  {journal} {\bibinfo  {journal}
  {Phys. Rev. Lett.}\ }\textbf {\bibinfo {volume} {10}},\ \bibinfo {pages}
  {159} (\bibinfo {year} {1963})}\BibitemShut {NoStop}%
\bibitem [{\citenamefont {Manmana}\ \emph {et~al.}(2007)\citenamefont
  {Manmana}, \citenamefont {Wessel}, \citenamefont {Noack},\ and\ \citenamefont
  {Muramatsu}}]{manma07}%
  \BibitemOpen
  \bibfield  {author} {\bibinfo {author} {\bibfnamefont {S.~R.}\ \bibnamefont
  {Manmana}}, \bibinfo {author} {\bibfnamefont {S.}~\bibnamefont {Wessel}},
  \bibinfo {author} {\bibfnamefont {R.~M.}\ \bibnamefont {Noack}}, \ and\
  \bibinfo {author} {\bibfnamefont {A.}~\bibnamefont {Muramatsu}},\ }\href@noop
  {} {\bibfield  {journal} {\bibinfo  {journal} {Phys. Rev. Lett.}\ }\textbf
  {\bibinfo {volume} {98}},\ \bibinfo {pages} {210405} (\bibinfo {year}
  {2007})}\BibitemShut {NoStop}%
\bibitem [{\citenamefont {Manmana}\ \emph {et~al.}(2009)\citenamefont
  {Manmana}, \citenamefont {Wessel}, \citenamefont {Noack},\ and\ \citenamefont
  {Muramatsu}}]{Manmana2009}%
  \BibitemOpen
  \bibfield  {author} {\bibinfo {author} {\bibfnamefont {S.~R.}\ \bibnamefont
  {Manmana}}, \bibinfo {author} {\bibfnamefont {S.}~\bibnamefont {Wessel}},
  \bibinfo {author} {\bibfnamefont {R.~M.}\ \bibnamefont {Noack}}, \ and\
  \bibinfo {author} {\bibfnamefont {A.}~\bibnamefont {Muramatsu}},\ }\href
  {\doibase 10.1103/PhysRevB.79.155104} {\bibfield  {journal} {\bibinfo
  {journal} {Phys. Rev. B}\ }\textbf {\bibinfo {volume} {79}},\ \bibinfo
  {pages} {155104} (\bibinfo {year} {2009})}\BibitemShut {NoStop}%
\bibitem [{\citenamefont {Barmettler}\ \emph {et~al.}(2009)\citenamefont
  {Barmettler}, \citenamefont {Punk}, \citenamefont {Gritsev}, \citenamefont
  {Demler},\ and\ \citenamefont {Altman}}]{Barmettler2009}%
  \BibitemOpen
  \bibfield  {author} {\bibinfo {author} {\bibfnamefont {P.}~\bibnamefont
  {Barmettler}}, \bibinfo {author} {\bibfnamefont {M.}~\bibnamefont {Punk}},
  \bibinfo {author} {\bibfnamefont {V.}~\bibnamefont {Gritsev}}, \bibinfo
  {author} {\bibfnamefont {E.}~\bibnamefont {Demler}}, \ and\ \bibinfo {author}
  {\bibfnamefont {E.}~\bibnamefont {Altman}},\ }\href {\doibase
  10.1103/PhysRevLett.102.130603} {\bibfield  {journal} {\bibinfo  {journal}
  {Phys. Rev. Lett.}\ }\textbf {\bibinfo {volume} {102}},\ \bibinfo {pages}
  {130603} (\bibinfo {year} {2009})}\BibitemShut {NoStop}%
\bibitem [{\citenamefont {Lieb}\ and\ \citenamefont
  {Robinson}(1972)}]{Lieb1972}%
  \BibitemOpen
  \bibfield  {author} {\bibinfo {author} {\bibfnamefont {E.~H.}\ \bibnamefont
  {Lieb}}\ and\ \bibinfo {author} {\bibfnamefont {D.~W.}\ \bibnamefont
  {Robinson}},\ }\href {\doibase 10.1007/BF01645779} {\bibfield  {journal}
  {\bibinfo  {journal} {Commun. Math. Phys.}\ }\textbf {\bibinfo {volume}
  {28}},\ \bibinfo {pages} {251} (\bibinfo {year} {1972})}\BibitemShut
  {NoStop}%
\bibitem [{\citenamefont {Jurcevic}\ \emph {et~al.}(2014)\citenamefont
  {Jurcevic}, \citenamefont {Lanyon}, \citenamefont {Hauke}, \citenamefont
  {Hempel}, \citenamefont {Zoller}, \citenamefont {Blatt},\ and\ \citenamefont
  {Roos}}]{Jurcevic2014a}%
  \BibitemOpen
  \bibfield  {author} {\bibinfo {author} {\bibfnamefont {P.}~\bibnamefont
  {Jurcevic}}, \bibinfo {author} {\bibfnamefont {B.~P.}\ \bibnamefont
  {Lanyon}}, \bibinfo {author} {\bibfnamefont {P.}~\bibnamefont {Hauke}},
  \bibinfo {author} {\bibfnamefont {C.}~\bibnamefont {Hempel}}, \bibinfo
  {author} {\bibfnamefont {P.}~\bibnamefont {Zoller}}, \bibinfo {author}
  {\bibfnamefont {R.}~\bibnamefont {Blatt}}, \ and\ \bibinfo {author}
  {\bibfnamefont {C.~F.}\ \bibnamefont {Roos}},\ }\href {\doibase
  10.1038/nature13461} {\bibfield  {journal} {\bibinfo  {journal} {Nature}\
  }\textbf {\bibinfo {volume} {511}},\ \bibinfo {pages} {202} (\bibinfo {year}
  {2014})}\BibitemShut {NoStop}%
\bibitem [{\citenamefont {Hauke}\ and\ \citenamefont
  {Tagliacozzo}(2013)}]{Hauke2013}%
  \BibitemOpen
  \bibfield  {author} {\bibinfo {author} {\bibfnamefont {P.}~\bibnamefont
  {Hauke}}\ and\ \bibinfo {author} {\bibfnamefont {L.}~\bibnamefont
  {Tagliacozzo}},\ }\href {\doibase 10.1103/PhysRevLett.111.207202} {\bibfield
  {journal} {\bibinfo  {journal} {Phys. Rev. Lett.}\ }\textbf {\bibinfo
  {volume} {111}},\ \bibinfo {pages} {207202} (\bibinfo {year}
  {2013})}\BibitemShut {NoStop}%
\bibitem [{\citenamefont {Eisert}\ \emph {et~al.}(2013)\citenamefont {Eisert},
  \citenamefont {van~den Worm}, \citenamefont {Manmana},\ and\ \citenamefont
  {Kastner}}]{Eisert2013}%
  \BibitemOpen
  \bibfield  {author} {\bibinfo {author} {\bibfnamefont {J.}~\bibnamefont
  {Eisert}}, \bibinfo {author} {\bibfnamefont {M.}~\bibnamefont {van~den
  Worm}}, \bibinfo {author} {\bibfnamefont {S.~R.}\ \bibnamefont {Manmana}}, \
  and\ \bibinfo {author} {\bibfnamefont {M.}~\bibnamefont {Kastner}},\ }\href
  {\doibase 10.1103/PhysRevLett.111.260401} {\bibfield  {journal} {\bibinfo
  {journal} {Phys. Rev. Lett.}\ }\textbf {\bibinfo {volume} {111}},\ \bibinfo
  {pages} {260401} (\bibinfo {year} {2013})}\BibitemShut {NoStop}%
\bibitem [{\citenamefont {J{\"{u}}nemann}\ \emph {et~al.}(2013)\citenamefont
  {J{\"{u}}nemann}, \citenamefont {Cadarso}, \citenamefont
  {P{\'{e}}rez-Garc{\'{i}}a}, \citenamefont {Bermudez},\ and\ \citenamefont
  {Garc{\'{i}}a-Ripoll}}]{Junemann2013}%
  \BibitemOpen
  \bibfield  {author} {\bibinfo {author} {\bibfnamefont {J.}~\bibnamefont
  {J{\"{u}}nemann}}, \bibinfo {author} {\bibfnamefont {A.}~\bibnamefont
  {Cadarso}}, \bibinfo {author} {\bibfnamefont {D.}~\bibnamefont
  {P{\'{e}}rez-Garc{\'{i}}a}}, \bibinfo {author} {\bibfnamefont
  {A.}~\bibnamefont {Bermudez}}, \ and\ \bibinfo {author} {\bibfnamefont
  {J.~J.}\ \bibnamefont {Garc{\'{i}}a-Ripoll}},\ }\href {\doibase
  10.1103/PhysRevLett.111.230404} {\bibfield  {journal} {\bibinfo  {journal}
  {Phys. Rev. Lett.}\ }\textbf {\bibinfo {volume} {111}},\ \bibinfo {pages}
  {230404} (\bibinfo {year} {2013})}\BibitemShut {NoStop}%
\bibitem [{\citenamefont {Wick}(1950)}]{Wick1950}%
  \BibitemOpen
  \bibfield  {author} {\bibinfo {author} {\bibfnamefont {G.~C.}\ \bibnamefont
  {Wick}},\ }\href {\doibase 10.1103/PhysRev.80.268} {\bibfield  {journal}
  {\bibinfo  {journal} {Phys. Rev.}\ }\textbf {\bibinfo {volume} {80}},\
  \bibinfo {pages} {268} (\bibinfo {year} {1950})}\BibitemShut {NoStop}%
\bibitem [{\citenamefont {Fioretto}\ and\ \citenamefont
  {Mussardo}(2010)}]{Fioretto2010}%
  \BibitemOpen
  \bibfield  {author} {\bibinfo {author} {\bibfnamefont {D.}~\bibnamefont
  {Fioretto}}\ and\ \bibinfo {author} {\bibfnamefont {G.}~\bibnamefont
  {Mussardo}},\ }\href {\doibase 10.1088/1367-2630/12/5/055015} {\bibfield
  {journal} {\bibinfo  {journal} {New J. Phys.}\ }\textbf {\bibinfo {volume}
  {12}},\ \bibinfo {pages} {055015} (\bibinfo {year} {2010})}\BibitemShut
  {NoStop}%
\bibitem [{\citenamefont {Lanczos}(1938)}]{lancz38}%
  \BibitemOpen
  \bibfield  {author} {\bibinfo {author} {\bibfnamefont {C.}~\bibnamefont
  {Lanczos}},\ }\href {\doibase 10.1002/sapm1938171123} {\bibfield  {journal}
  {\bibinfo  {journal} {Studies in Applied Mathematics}\ }\textbf {\bibinfo
  {volume} {17}},\ \bibinfo {pages} {123} (\bibinfo {year} {1938})}\BibitemShut
  {NoStop}%
\bibitem [{\citenamefont {Arnoldi}(1951)}]{Arnoldi1951}%
  \BibitemOpen
  \bibfield  {author} {\bibinfo {author} {\bibfnamefont {W.~E.}\ \bibnamefont
  {Arnoldi}},\ }\href {\doibase 10.1090/qam/42792} {\bibfield  {journal}
  {\bibinfo  {journal} {Quarterly Appl. Math.}\ }\textbf {\bibinfo {volume}
  {9}},\ \bibinfo {pages} {17} (\bibinfo {year} {1951})}\BibitemShut {NoStop}%
\bibitem [{\citenamefont {Damascelli}, \citenamefont {Shen},\ and\
  \citenamefont {Hussain}(2003)}]{damas03}%
  \BibitemOpen
  \bibfield  {author} {\bibinfo {author} {\bibfnamefont {A.}~\bibnamefont
  {Damascelli}}, \bibinfo {author} {\bibfnamefont {Z.-X.}\ \bibnamefont
  {Shen}}, \ and\ \bibinfo {author} {\bibfnamefont {Z.}~\bibnamefont
  {Hussain}},\ }\href@noop {} {\bibfield  {journal} {\bibinfo  {journal} {Rev.
  Mod. Phys.}\ }\textbf {\bibinfo {volume} {75}},\ \bibinfo {pages} {473}
  (\bibinfo {year} {2003})}\BibitemShut {NoStop}%
\bibitem [{\citenamefont {Bloch}, \citenamefont {Dalibard},\ and\ \citenamefont
  {Zwerger}(2008)}]{bloch08}%
  \BibitemOpen
  \bibfield  {author} {\bibinfo {author} {\bibfnamefont {I.}~\bibnamefont
  {Bloch}}, \bibinfo {author} {\bibfnamefont {J.}~\bibnamefont {Dalibard}}, \
  and\ \bibinfo {author} {\bibfnamefont {W.}~\bibnamefont {Zwerger}},\
  }\href@noop {} {\bibfield  {journal} {\bibinfo  {journal} {Rev. Mod. Phys.}\
  }\textbf {\bibinfo {volume} {80}},\ \bibinfo {pages} {885} (\bibinfo {year}
  {2008})}\BibitemShut {NoStop}%
\bibitem [{\citenamefont {Hamerla}(2013)}]{Hamerla2013b}%
  \BibitemOpen
  \bibfield  {author} {\bibinfo {author} {\bibfnamefont {S.~A.}\ \bibnamefont
  {Hamerla}},\ }\emph {\bibinfo {title} {{Dynamics of Fermionic Hubbard Models
  after Interaction Quenches in One and Two Dimensions}}},\ \href@noop {}
  {\bibinfo {type} {Phd thesis}},\ \bibinfo  {school} {Technische
  Universit{\"{a}}t Dortmund} (\bibinfo {year} {2013})\BibitemShut {NoStop}%
\end{thebibliography}

%

\end{document}